\documentclass[pre,preprint]{revtex4}
\usepackage{epsfig}
\usepackage{amsfonts,graphicx,bbm,color,amssymb,amsmath}

\begin{document}

\title{Dichotomous Markov Noise: \\Exact results for out-of-equilibrium systems\\
(a brief overview)}

\date{\today}

\author{Ioana Bena}
\affiliation{Theoretical Physics Department, University of Geneva,\\
 Quai Ernest Ansermet no.24, CH-1211 Geneva 4, Switzerland\\
Ioana.Bena@physics.unige.ch}

\begin{abstract}
Nonequilibrium systems driven by additive or multiplicative
dichotomous Markov noise  appear in a wide variety
of physical and mathematical models.
We review here some prototypical examples,
with an emphasis on {\em analytically-solvable} situations.
In particular, it has escaped attention till
recently that the standard results for the
long-time properties of such systems cannot
be applied when unstable fixed points are crossed
in the asymptotic regime. We show how calculations
have to be modified to deal with these cases and
present a few relevant applications  -- the hypersensitive
transport,  the rocking ratchet, and the stochastic
Stokes' drift.
These results reinforce the impression that dichotomous
noise can be put on a par with Gaussian white noise as
far as obtaining analytical results is concerned.
They convincingly illustrate the interplay between noise
and nonlinearity in generating nontrivial behaviors of
nonequilibrium systems and point to various practical applications.
\end{abstract}

\maketitle

\section{INTRODUCTION}

The statistical
theory of out-of-equilibrium systems is one of the
most challenging, rapidly evolving, and an interdisciplinary
domain of modern research. Its fundamental importance is rooted in the
fact that the majority of the situations encountered in
nature  (in its broadest sense, including physical, chemical,
and biological systems) are nonequilibrium ones. Such
systems exhibit very complex and often counter-intuitive
behaviors, resulting from a
generic interplay between a large number of degrees of
freedom, nonlinearities, ``noise" or perturbations of various origins,
driving forces and dissipation (corresponding to various levels of
coarse-grained description).

Usually,  noise and  stochastic equations appear in the modeling
when one concentrates the study
on a few relevant variables
(hereafter called  ``system under study") and  approximates
the effects of the eliminated  degrees of freedom
through a ``random  force" or ``noise" with prescribed statistical properties.
In particular, these eliminated degrees of freedom have a characteristic time scale,
that is translated into the specific correlation time of the noise that
intends to mimic them.

In many situations of interest,
the characteristic response time of the system under study
is much larger than this specific time scale of the
eliminated variables.
Following the seminal works of Einstein, Langevin
and Smoluchowsky, the noise is then safely modeled
as a Gaussian white noise  (GWN) that is $\delta$-correlated in time.
In this case the
model system  is generically referred to as
a ``Brownian particle". The Brownian motion under the action of the
GWN  is a Wiener  process, and
a detailed mathematical analysis can be carried on the basis of
Langevin or Fokker-Planck equations.

The Wiener process and the GWN are, of course,  stochastic processes
of fundamental importance.
However, they do not exhaust all the situations one may be
called to model. Indeed, there are cases when the eliminated degrees of freedom are {\em slow} on the time scale of the studied system. Then one
has to mimic the result of the coarse-graining over such a set of slow variables
through a {\em colored noise}, i.e., a noise that has
a non-negligible correlation time (and thus a non-flat power spectrum).
Such noises have been studied in great detail in zero-dimensional
systems, and their specific properties are known to have a profound
influence on the behavior of these systems.
Generically, they lead the system out of equilibrium
(they break the detailed balance in the configuration space of the
system).
The effect of the color of the noise on noise-induced transitions
and phase transitions
continues to be documented, and has been found to be quite important,
e.g., it can alter the type of transition and lead to re-entrance
phenomena. Other noise-induced effects have also been found to be
sensitive to the correlation time of the noise, like the stochastic resonance,
the synchronization of several noisy dynamical units, and the
directed transport in ratchets.

The two most commonly discussed examples of colored noise are the
Ornstein-Uhlenbeck process and the dichotomous Markov noise (DMN).
Although the Ornstein-Uhlenbeck process is easily invoked
in view of Doob's theorem~\footnote{Roughly speaking, Doob's theorem states that the Ornstein-Uhlenbeck process is the only stationary Gaussian diffusion process. See, e.g., 
Ref. ~\cite{vankampen92}  for more details.},
the purpose of this review is to show that {\em DMN has
particular virtues and interest.}

But before proceeding with this argumentation, 
let us define the DMN and describe its
main stochastic properties.

\subsection{Definition of DMN}
\begin{figure}[h!]
\quad \vspace{1.5cm}\\
\centerline{\psfig{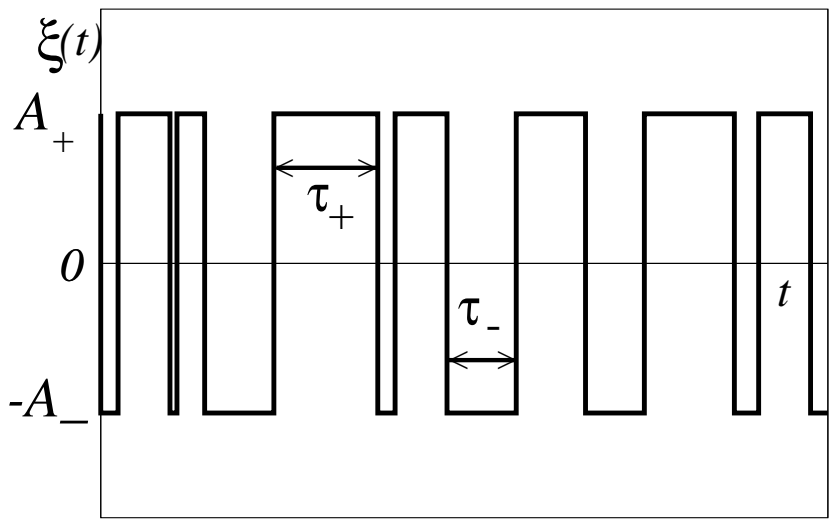}} \vspace*{8pt}
\caption{A realization of an asymmetric DMN $\xi(t)$ that jumps between two values $\pm A_{\pm}$ with constant transition rates $k_{\pm}$. The waiting times $\tau_{\pm}$
in the two states are exponentially-distributed stochastic variables,
thus ensuring the Markovian character of the DMN, see the main text. }
\label{figure1}
\end{figure}

The DMN  $\xi(t)$ is a very simple {\em two-valued $\pm A_{\pm}$ stochastic process, with constant transition rates $k_{\pm}$ between the two states}~\cite{vankampen92,horsthemke84}. Figure~\ref{figure1}
illustrates a realization of DMN.
The constancy of the transition
rates corresponds to exponentially distributed waiting times $\tau_{\pm}$ in the two states,
\begin{equation}
\mbox{Probability}(\tau_{\pm})=k_{\pm}\exp(-k_{\pm}\tau_{\pm})
\end{equation}
(i.e., the transitions are driven by Poisson renewal processes) and DMN is  {\em Markovian}. It is therefore completely characterized by the initial state and the
matrix of the  transition probabilities,
\begin{eqnarray}
\left[P_{ij}(t)\right]_{i,j=\pm}&=&\left[\mbox{Probability}(\xi(t)=iA_i|\xi(0)=jA_j)\right]_{i,j=\pm}
\nonumber\\
&&\nonumber\\
&=&\left[
\begin{array}{cc}
P_{--}(t) &\;\;\; P_{-+}(t)\\
P_{+-}(t) &\;\;\; P_{++}(t)
\end{array}
\right]=
\tau_c
\left[
\begin{array}{cc}
k_++k_-e^{-t/\tau_c} &\;\;\; k_+(1-e^{-t/\tau_c})\\
k_-(1-e^{-t/\tau_c}) &\;\;\; k_-+k_+e^{-t/\tau_c}
\end{array}
\right]\,,
\end{eqnarray}
where
\begin{equation}
\tau_c=\frac{1}{k_++k_-}
\end{equation}
is the characteristic relaxation time to the stationary state of the DMN.
In the foregoing we shall be exclusively concerned with {\em stationary DMN},  for which the stationary probabilities of the two states are
\begin{equation}
\mbox{Probability}(\xi=A_+)={k_-}\,\tau_c\;,\;\;
\mbox{Probability}(\xi=-A_-)=k_+\,\tau_c\,,
\end{equation}
with the corresponding mean value
\begin{equation}
\langle\xi(t)\rangle =(k_-A_+-k_+A_-)\,\tau_c\,.
\end{equation}
Moreover, we shall generally consider zero-mean DMN, $\langle\xi(t)\rangle=0$,
in order to avoid any systematic bias introduced by the noise in the dynamics of the driven system.
The stationary temporal autocorrelation function of DMN
is exponentially-decaying
\begin{equation}
\langle \xi(t) \xi(t')\rangle =
\displaystyle\frac{D}{\tau_c}\exp\left(-\frac{|t-t'|}{\tau_c}\right)\,,
\end{equation}
corresponding to a {\em finite correlation time} $\tau_c$ and an ``amplitude"
\begin{equation}
D=k_+k_-\tau_c^3(A_++A_-)^2\,.
\end{equation}
The power spectrum is thus a Lorenzian related to the characteristic time scale $\tau_c$,
\begin{equation}
S(\omega)=\frac{D}{\pi(1+\omega^2\,\tau_c^2)}\,,
\end{equation}
so the DMN is indeed {\em colored}.

A particular case of DMN, that will be widely used in the foregoing, is that of the
{\em symmetric} DMN, for which $A_+=A_-\equiv A$ and $k_+=k_-\equiv k$,
so that the correlation time is $\tau_c=(2k)^{-1}$, and the ``amplitude" 
$D=A^2/(2k)$, with, moreover, $\langle\xi(t)\rangle=0$.

\begin{figure}[h!]
\quad \vspace{1.5cm}\\
\centerline{\psfig{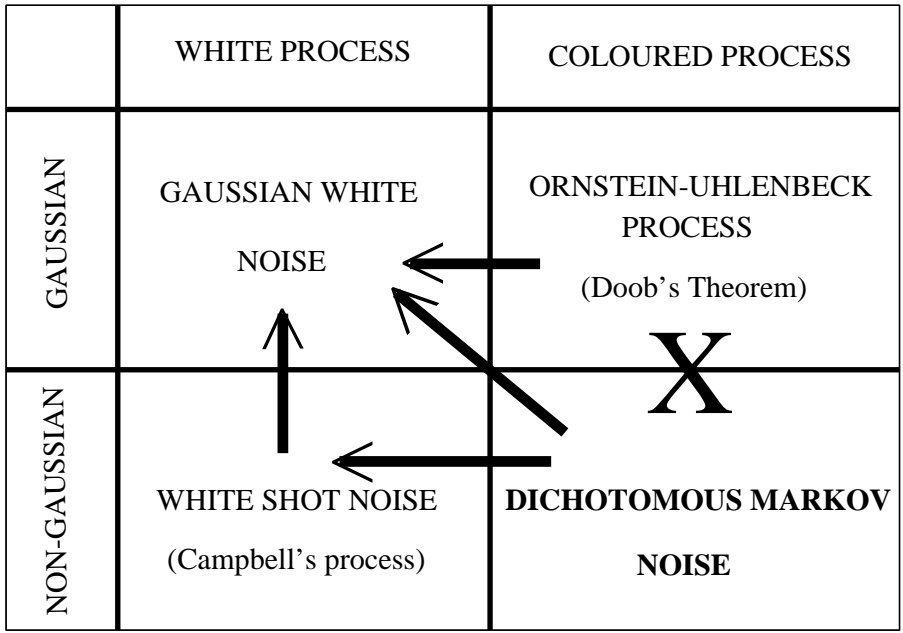}} \vspace*{8pt}
\caption{Schematic representation of the relationships between DMN and other important stationary, zero-mean stochastic processes. DMN reduces to white shot noise or to Gaussian white noise in the appropriate limits (see the main text); however, DMN and the Ornstein-Uhlenbeck process cannot be mapped one onto the other.  }
\label{figure2}
\end{figure}

Let us underline yet another important property of the stationary,
zero-mean DMN, namely that in  appropriate limits it  reduces either
to a {\em white shot noise} (WSN), or to a {\em Gaussian white
noise} (GWN), see Ref.~\cite{vandenbroeck83}  for a detailed
discussion. Figure~\ref{figure2} represents schematically these
relationships.

Consider the stationary {\em asymmetric} DMN of zero mean,
characterized by {\em three independent parameters} (e.g., $k_{\pm}$ and $A_+$),
with $A_+/k_+=A_-/k_-\equiv \lambda$. Then, by taking the limit
\begin{equation}
A_-,\,k_-\rightarrow \infty\,,\;\mbox{while keeping} \;\;A_-/k_-=\lambda=\mbox{finite}\,,
\end{equation}
one recovers the WSN driven by a Poisson process  (also called the Campbell's process);
this one is characterized by {\em two parameters} ($\lambda$ and, e.g., $k_+$). Considering further on the limit $A_{+}, \,k_{+}\rightarrow \infty$ with $\lambda \rightarrow 0$ but $D=\lambda^2{k_+}=A_+^2/k_+=\mbox{finite}$,
WSN reduces to a GWN; this latter is characterized by a {\em single parameter}, the noise ``amplitude" $D$.

Note also that the {\em symmetric} DMN can reduce directly (i.e., without passing through the stage of a WSN) to the GWN by taking simultaneously the limits $A,\,k\rightarrow \infty$, such that $D=A^2 \,\tau_c=A^2/(2k)=\mbox{finite}$ represents the amplitude of the GWN.

There is no mapping possible between DMN and the Ornstein-Uhlenbeck process:
indeed, one cannot eliminate the non-Gaussian character of the DMN while keeping a finite correlation time. The two noises have distinct statistical properties, a point that is
also generally reflected in the response properties of the dynamical systems
driven by these noises. However, as indicated by the central limit theorem, the superposition of $M$ suitably-scaled, independent DMNs $\xi_i(t)\,,i=1,...,M$\,,
converges in the limit $M\rightarrow \infty$ to an Ornstein-Uhlenbeck process.
This property may be used to tackle systems driven by an Ornstein-Uhlenbeck noise
(that are notoriously difficult to study)
by constructing approximate solutions corresponding to the superposition  of
several DMNs.

\subsection{Motivation, relevance, and the importance of using DMN}

This review is intended to present DMN as a  {\em tool for modeling stochastic processes}.
We shall thus try to clarify its applicability, flexibility, and limits, as well as to describe a few prototypical applications. Several essential points,
briefly resumed below,
will be resulting from this review.
\begin{itemize}
\item As it will become clear, systems driven by DMN are {\em encountered in a wide variety of physical and mathematical models}.  Besides the ``fashion-effect" (peaked around the 1970s), there are deeper reasons to this and we shall  address them.\\
\item A  basic point is that {\em DMN mimics the effects of
finite correlation time (colour) of the noise  in a very simple way}. It constitutes thus a good alternative to the widely-spread Ornstein-Uhlenbeck process, which is often quite untractable analytically. So, the DMN reflects the effect of the eliminated {\em slow} degrees
of freedom on the dynamics of the relevant variables under study. Refer for example to  Refs.~\cite{hanggi95,vankampen76,vankampen85,zwanzig01}
for detailed discussions on this essential point of the modeling of stochastic systems.

Moreover, the interplay between the {\em intrinsic} time scale of the DMN and other time scales present in the problem (e.g., characteristic relaxation time of the system under study, external periodic perturbations, etc.)  may lead to nontrivial effects (e.g., multistability and hysteretic behavior, stochastic resonance, synchronization effects, etc.). These effects are
absent when a white noise is acting (even multiplicatively !) on the system.\\

\item One has to realize that {\em DMN may be directly a very good representation
of a simple and  frequently-encountered physical situation},
namely that of the {\em  thermally activated transitions between two configurations/states} of a system
(as far as the ``intra-configuration" motion is unimportant,  filtered-out or coarse-grained).

A  very clear illustration is offered by the  small electronic devices (e.g., MOSFET, MIM, JFET, point-contact resistances, etc.), as described in the review article  Ref.~\cite{kirton89}. The active volume of such devices is so small that they contain only a reduced number of charge carriers; then the alternate capture and emission of carriers at individual defect sites generates {\em measurable} discrete jumps in the device resistance. These jumps
are naturally modeled as a DMN process. Their study provides a powerful mean of investigating the nature of defects at the interfaces, the kinetics of carriers' capture and emission (see, e.g., Ref.~\cite{xiao04} for a  recent experimental  example),
and has demonstrated the defect origin of low-frequency $1/f$ noise in these devices, etc.\\

\item Generically,  DMN {\em drives the system out-of-equilibrium} (i.e., it breaks the detailed balance in the configuration space of the system). Therefore it may lead to novel behaviors that are not accessible in equilibrium systems (like, for example, directed transport out of microscopic fluctuations). In particular, memory effects may become important -- the driven system is typically non-Markovian.\\

\item On one hand, DMN is ``sufficiently simple" so that it {\em often allows for full
analytical description}, and it represents therefore a good research, as well as didactical tool. The emphasis of our presentation will thus be, to the largest possible extent, on {\em analytically-solvable} models and on {\em exact results},
as stated in the title of this review~\footnote{One should keep in mind the difficulties generically encountered in the analysis of out-of-equilibrium systems (when very
often one has to appeal to various approximations),
and therefore the scarcity of exact results for such systems.}.

Moreover, the ``simplicity" of DMN allows to dissect the essential mechanisms 
 at work behind various nonequilibrium, 
 often seemingly complex phenomena.  One can 
thus  reveal the ``minimal scenario" and  ingredients 
for the appearance of these processes.\\

\item On the other hand, DMN is ``sufficiently rich" so that it leads to {\em highly nontrivial and varied nonequilibrium behaviors in nonlinear systems}. This illustrates once more the fundamental fact that stochasticity may affect the dynamics more strongly than just a simple perturbation around the deterministic evolution: it may induce {\em qualitatively} different behaviors.\\

\item As it was mentioned above, DMN {\em reduces, in appropriate limits, to WSN and to GWN}, and thus offers an alternative way of solving problems connected with these noises (in particular, master equation for DMN is often simpler to solve than the Fokker-Planck equation for GWN).

Let us also note that the relationship of DMN with the GWN
played a role in the famous (and unfounded~!)
{\em It\^o versus Stratonovich  dilemma}~\footnote{Recall that a stochastic differential equation with {\em multiplicative} GWN is not defined unambiguously: one has to supply it with an integration rule. Two such rules were essentially proposed in the existing literature, namely the {\em It\^o interpretation} (and the use of martingale formalism), and the {\em Stratonovich interpretation}.  The controversy arose as to which of these two conventions was ``the correct one".  The fact that the master equation for
a system driven by DMN (for which there is no ambiguity in interpretation) reduces,
in the appropriate limit,  to the Stratonovich form of the Fokker-Planck equation for the GWN was used as an argument in favor of the Stratonovich interpretation. However, as explained in detail in Ref.~\cite{vankampen81}, such a controversy is actually completely meaningless.}, see
Refs.~\cite{vankampen81}  and \cite{vandenbroeck83}  for pertinent comments on this point. \\

\item From the point of view of numerical simulations, DMN has the advantage of
{\em being easy to implement as an external noise with finite support}.\\

\item Last, but surely not least,
systems driven by DMN point to {\em interesting practical applications}.
\end{itemize}

\subsection{The structure of the review}

Despite an impressive literature on the systems driven by DMN, 
the main lines and subjects
of interest are easy to infer. The type of systems that are generically studied are {\em zero-dimensional dichotomous flows}, corresponding to a single stochastic variable driven by a DMN that may be additive or multiplicative, as described in Sec.~II.
We shall  be first addressing in Sec.~III
the problem of the transient (time-dependent) characteristics of such flows, that might be important for some practical applications and/or finite observation times.
We shall classify the ``solvable" cases and  discuss in detail the seminal example of the dichotomous  diffusion on a line. We shall turn afterwards to the ``more productive" study of the
asymptotic, long-time behavior of the dichotomous flows. As it will become clear, both the physics and the mathematical approach are very different if the asymptotic dynamics exhibits or not {\em unstable critical points} -- this represents  a main point of this review. The ``standard old theories" were limited to the absence of unstable critical points, as described in Sec.~IV. Most of the results referred to the celebrated noise-induced transitions and phase transitions that we shall briefly present. In Sec.~V we turn to the situations when the asymptotic dynamics presents unstable critical points. We show how calculations have to be modified in order to deal with such situations
and illustrate the method on three prototypical examples, namely the hypersensitive response, the rocking ratchet, and the stochastic Stokes' drift. Section~VI will be devoted to a brief presentation of escape statistics (mean first-passage time problems, resonant activation over a fluctuating barrier, etc.). In Sec.~VII we shall  discuss  the stochastic resonance phenomenon
and its DMN-induced enhancement.
Section~VIII is devoted to some comments on spatial patterns induced by DMN forcing, while Sec.~IX describes briefly random maps with DMN. A few conclusions and perspectives are relegated to Sec.~X.

A last warning refers to the fact that this review is non-exhaustive
(and we do apologize for the omissions), pedagogical to a large extent, and rather non-technical. We emphasized subjects that count amongst the most 
recent in this domain and in which we were directly involved. 
But, of course, all important earlier results are also 
described in detail, for the sake of a correct perspective 
on the field. 

\section{SYSTEMS DRIVEN BY DMN: DEFINITION OF DICHOTOMOUS FLOWS}

A  zero-dimensional  dichotomous flow corresponds  to the temporal evolution of some characteristic scalar variable $x=x(t)$ of the system under study, whose velocity switches at random  between two dynamics:  the ``+" dynamics, $\dot{x}(t)=f_+(x)$, and the ``--" dynamics,  $\dot{x}(t)=f_-(x)$ (the dot designates the time-derivative). This process can be described by the following stochastic differential equation:
\begin{equation}
\dot{x}(t)=f(x)\,+\,g(x)\,\xi(t)\,,
\end{equation}
where $\xi(t)$ is a realization of the DMN taking the values $\pm A_{\pm}$
with transition rates $k_{\pm}$ between these values, and $f_{\pm}(x)=f(x)\pm g(x)A_{\pm}$.
If $g(x)$ is a constant, the DMN acts {\em additively}; otherwise, for $g(x)\neq$ constant,
the noise is {\em multiplicative}.
 We shall consider throughout the paper only {\em constant} transition rates
$k_{\pm}$, although some recent  stochastic nonequilibrium models of protein motors
use $x$-dependent transition rates $k_{\pm}(x)$, see Ref.~\cite{julicher99}.
Most of our results can be
easily generalized to cover these situations too.  Moreover, if not stated explicitely otherwise, we shall be working with a {\em symmetric DMN}, $A_{\pm}\equiv A$ and $k_{\pm}\equiv k$.

For the simplicity of the presentation, we shall often refer to $x(t)$ as the  
``position of an overdamped
particle"; but one should keep in mind that its actual nature depends on the system under study (i.e., that $x(t)$ can be a spatial coordinate, a current, a concentration, a reaction coordinate, etc.).
It is a stochastic process, and for most of the practical purposes its properties  can be essentially described
through the properties and evolution equation of the
probability distribution function $P(x,t)$~\footnote{See, e.g., Ref.~\cite{kubo}
for the discussion of the complete characterization of a stochastic process in terms of various associated probability distribution functions.}.
Indeed, one is in general interested by the
the {\em mean over the realizations of the DMN} of an arbitrary function ${\cal F}(x)$ of the stochastic
variable $x(t)$, a type of quantity that we shall denote throughout the text  by $\langle ... \rangle$: \begin{equation}
\langle {\cal F} \rangle =\int dx P(x,t) {\cal F}(x) \,.
\end{equation}
In the nonstationary regime of the dichotomous flow this is an explicit function of time, as discussed below.

\section{THE TIME-DEPENDENT PROBLEM}

One is, of course,  tempted to address first the question of the non-stationary, transient stochastic behavior of the
variable $x(t)$, i.e., that of the temporal evolution of its probability density from a given initial state
(a given initial probability distribution function) to the
(presumably existent) asymptotic stationary state with the corresponding stationary
probability distribution function.

\subsection{The master equation and the time-dependent ``solvable" cases}

The master equation for the probability density of the compound (or vectorial) stochastic process $[x(t),\,\xi(t)]$,  namely $P_{\pm}(x,t)=P(x,t;\xi=\pm 1)$, can be easily written down. It corresponds to
a Liouville flow in the phase space (describing  the deterministic evolution between two jumps of the DMN) plus a gain-and-loss term (reflecting the switch of the DMN between its two values):
\begin{eqnarray}
\partial_tP_+(x,t)&=&-\partial_x[f_+(x)\;P_+]-k(P_+-P_-)\,, \\
\partial_tP_-(x,t)&=&-\partial_x[f_-(x)\;P_-]+k(P_+-P_-)
\end{eqnarray}
(see, e.g., Ref.~\cite{balakrishnan93} for a detailed derivation).
One is interested, however, in the
evolution equation for the marginal probability
density of the stochastic variable $x(t)$, ${P(x,\,t)=P_+(x,t)+P_-(x,t)}$, for which one
obtains (using the above equations):
\begin{eqnarray}
\partial_t  {P(x,t)}&&=-\partial_x[f(x){P(x,t)}]\nonumber\\
&&+ A^2\,
\partial_x \,g(x)\int_0^t dt_1\exp[(-\partial_x
f(x)-2k)(t-t_1)]\,
\partial_x[g(x){P(x,t_1)}]\,.
\label{master}
\end{eqnarray}
This is an  intricate {\em integro-differential} equation
in time, with a kernel that involves the exponential of the differential operator $\partial_x$.
Thus, although the DMN is such a simple stochastic process,
the statistics of the driven process $x(t)$ may be remarkably complicated
and, in general,  $x(t)$ {\em is not a Markovian process}.
While the stationary solution
$P_{st}(x)$ to which the probability density $P(x,t)$ tends in the asymptotic limit $t\rightarrow \infty$ can be computed under rather general conditions (see Secs.~IV and V below), one can also address the legitimate question of the solvability of this
time-dependent master equation (\ref{master}).

A particularly relevant problem
 is whether one could construct/recover some Markovian process out of $x(t)$.  As shown
recently in Refs.~\cite{balakrishnan01I} and \cite{balakrishnan03}, this is achieved in the so-called {\em solvable cases}, when $P(x,t)$  obeys a closed partial differential equation
of finite-order in time.
If the order of this equation is $n \geqslant 1$, then one can construct a finite-dimensional vectorial Markovian process out of $x(t)$ and its temporal derivatives; more precisely,
 the vectorial process $\left[x(t), \, \dot{x}(t),\, ... , x^{(n-1)}(t)\right]$ is Markovian.

It is in this very point that lies the importance of knowing whether
a dichotomous flow is solvable:
if  one can reconstruct  a Markovian property at some level
of differentiation of $x(t)$,  then the knowledge of a finite-number
of initial conditions for the probability density and its time-derivatives
are enough in order to determine entirely the  subsequent evolution of $P(x,t)$.

 As discussed in detail in Refs.~\cite{balakrishnan01I} and
 \cite{balakrishnan03}, the
 {\em condition of solvability} is related to the behavior of the differential operators
 \begin{equation}
{{\cal A}}=-\partial_x[f(x)\,...]\;,\quad { {\cal B}}=-\partial_x[g(x)\,...]
\end{equation}
and the hierarchy of their comutators,
\begin{equation}
{{\cal C}_n}=\left[{ {\cal A}},\,{{\cal C}_{n-1}}\right]\quad (n\geqslant 1)\;,\quad \mbox{with}\; \quad{ {\cal C}_0}={{\cal B}}\,.
\end{equation}
If this hierarchy closes, then $P(x,t)$ satisfies a finite-order differential equation in time~\footnote{The justification of these conditions is rather long and delicate, and will not be given  here
(it involves the use of the stochastic Liouville equation for the phase-space density, Van Kampen's lemma,  the Shapiro-Longinov~\cite{shapiro} formula of differentiation  for a functional of the noise $\xi(t)$, etc.).  See Ref.~\cite{balakrishnan01I} for the details of their derivation. }.
In more detail, if
\begin{equation}
{{\cal C}_{{n}}}=\displaystyle\sum_{k=0}^
{{{ n-1}}}\beta_k\,{ {\cal C}_k}\,,
\end{equation}
with $\beta_k$ some constants, then $P(x,t)$ satisfies a partial differential equation of order
$(n+1)$ in time. If the linear combination involves ${\cal A}$ as well, i.e., if
\begin{equation}
{ {\cal C}_{{n}}}={\alpha
{ {\cal A}}}+\displaystyle\sum_{k=0}^{{{ n-1}}}\beta_k\,{ {\cal C}_k}\,,
\end{equation}
then the order of the equation is $(n+2)$.

Despite the seeming simplicity of  these criteria, one should realize that
they are quite restrictive and thus the classes of  solvable cases are rather reduced.

In particular,  the following cases are {\em not solvable}:

(a) The cases with an additive DMN ($g(x)=$ constant) and a nonlinear drift
term ($f(x)=$ nonlinear in $x$).

(b) When either $f(x)$ or $g(x)$ is a polynomial of order $\geqslant 2$, $P(x,t)$ does not satisfy {\em in general} a finite-order equation (although it may do so in special cases of specific relationships between the coefficients of the polynomials).

A few examples of {\em solvable cases}:

(a) When $f(x)=\mbox{constant}\cdot g(x)$, i.e., ${\cal C}_1=0$, the process $x(t)$ can be mapped, through a nonlinear transformation, onto a pure {\em dichotomous diffusion}
for which the time-dependent solution is well-known (see the subsection below).

(a1) A subcase of interest is that of $f(x)=Ag(x)$, so that $f_-(x)=0$. This corresponds to the so-called delayed evolution, in which the deterministic dynamics
governed by the flow $f_+(x)=2f(x)$ is interrupted at random instants and $x$ remains frozen at its current value, till the noise switches back  and the
``$+$"  dynamics is continued.

A closely-related type of flow, called interrupted evolution, is characterized by the fact that in the quiescent ``--" state  ($f_-=0$) the stochastic variable is reset
to a random value drawn from a fixed distribution. See Ref.~\cite{balakrishnan01II} for more details and applications of these two types of flows.

(a2) When $f(x)=0$, i.e., $f_+(x)=-f_-(x)$, the problem reduces again to the dichotomous diffusion. This case is of interest, e.g.,  in problems involving the exchange of stability between two critical points of the alternate dynamics, see Ref.~\cite{balakrishnan01II}.

(b) Another case presented in the literature, see Ref.~\cite{sancho84}, is that of ${\cal C}_1=-\beta_0{\cal B}$,
i.e.,  $f'(x)g(x)-g'(x)f(x)=\beta_0g(x)$, when $P(x,t)$ obeys a second-order differential equation. The early-day Hongler's model, see Ref.~\cite{hongler79},
with $f(x)=-\mbox{tanh}(x)$ and $g(x)=\mbox{sech}(x)$ fall in this class.

An important warning, however:  one should be aware of the fact that  ``solvable" in the sense
indicated here above does not imply, in general, that one can express $P(x,t)$ in a simple, closed algebraic form of some ``standard" functions. Indeed, as explained in Refs.~\cite{balakrishnan01I} and \cite{balakrishnan03}, this is merely an exceptional situation, and probably the only case that was completely explored till now is that of the dichotomous diffusion
(and processes that reduce to it through some change of variables).

\subsection{Dichotomous diffusion on a line}

It describes the  stochastic position $x(t)$ of a particle whose velocity
is a DMN,
i.e., it is  represented by the following stochastic differential equation:
\begin{equation}
\dot{x}(t)=\xi(t)\,.
\end{equation}
The corresponding master equation for the probability densities $P_{\pm}(x,t)$ is
\begin{equation}
\partial_t P_{+}=-A\partial_xP_++k(P_--P_+)\,,\quad
\partial_t P_{-}=A\partial_xP_-+k(P_+-P_-)
\label{ppm}
\end{equation}
(for the case of a symmetric DMN $\pm A$ with transition rate $k$).

This simple process $x(t)$ is an example of so-called {\em persistent diffusion
on a line} as detailed in Refs.~\cite{furth20,taylor21,goldstein51,balakrishnan88I}. Indeed, it can be
obtained~\footnote{In the same way as the normal diffusion is
obtained from the usual, simple discrete random walk on a lattice.} as the continuum limit of a ``persistent" random walk on a
$1D$ lattice, i.e., a random walk for which the transition probabilities left or right
at a given stage depend on the {\em direction} of the preceding jump. This means that the jump probabilities have a ``memory" of the previous state of the system, and therefore $x(t)$ is no longer Markovian. However, according to the general discussion in Sec.~III.A,  the vectorial process $[x(t), \,\dot{x}(t)]$ is Markovian. The probability distribution $P(x,t)$ obeys a second-order (in time) hyperbolic partial differential equation known as the {\em telegrapher's equation}~\footnote{This equation is well-known in the theory of electromagnetic signal propagation in transmission lines, and describes both the ``ballistic" transmission and the damping (diffusion) of the signal, see the main text. A model with $x$-dependent transition rates $k=k(x)$ corresponds to inhomogeneities in the transmission cables, see Ref.~\cite{hongler86}. }:
\begin{equation}
(\partial_{tt}+2k\partial_t-A^2\partial_{xx})P(x,t)=0\,.
\label{telegrapher}
\end{equation}
Note that each of $P_{\pm}(x,t)$ also obeys this equation.
Telegrapher's equation can be solved for various initial conditions of $P(x,t)$ and its time-derivative $\dot{P}(x,t)$~\cite{balakrishnan88I}.
The short-time behavior  (for $t \ll \tau_c=1/2k$) is governed by
the wave-like part of the equation, i.e., the mean square
displacement behaves ballistically: $\langle x^2(t)\rangle \approx A^2t^2$;
while the long-time behavior (for $t \gg \tau_c=1/2k$) is of course diffusive,
$\langle x^2(t)\rangle \approx (A^2/k)t=2Dt$, in
agreement with the central-limit theorem.

One can easily illustrate this behavior on the example of the ``symmetric" initial conditions
 $P(x,t=0)=\delta(x)$ and $\dot{P}(x,t=0)=0$, when the solution
 of the telegrapher's equation can be expressed in terms of  modified Bessel functions of zeroth and first order as:
\begin{eqnarray}
P(x,t)&=&\frac{e^{-kt}}{2}[\delta(x-At)+\delta(x+At)]+
\frac{ke^{-kt}}{2A}\left[\frac{}{}I_0\left(k\sqrt{t^2-x^2/A^2}\right)\right.\nonumber\\
&&\left.+\frac{t}{\sqrt{t^2-x^2/A^2}}I_1\left(k\sqrt{t^2-x^2/A^2}\right)\right]\;\left[\theta(x+At)-\theta(x-At)\right]\,.
\end{eqnarray}
The terms in $\delta$ (of amplitude that is decreasing exponentially in time)
describe the ballistic motion corresponding to the persistence of DMN in the
``+" or ``--" state during time $t$. The remaining terms represent the sum of contributions resulting from trajectories that imply one, two, etc ... transitions between the ``+" and ``--" states of the DMN during time $t$, and tend to a Gaussian
(reflecting an usual Brownian motion) in the asymptotic limit. Henceforth,
\begin{equation}
\langle x^2(t) \rangle =
\frac{A^2}{2k^2}\,\left[2kt-1+\exp(-2kt)\right]\,,
\end{equation}
with the mentioned ballistic, respectively diffusive behavior for short and long times.

\subsubsection{Applications}

As reviewed in Ref.~\cite{vandenbroeck90},
several  physical situations can be modeled through a dichotomous diffusion
and the telegrapher's equation, and we present them briefly below.
For most of them the transient regime and its stochastic properties are important, in view of the
finite-observation time, or the limited dimensions of the experimental device;
from here the relevance of knowing the time-dependent statistics, and not only the asymptotic regime.

\begin{figure}[thb]
\quad \vspace{1.5cm}\\
\centerline{\psfig{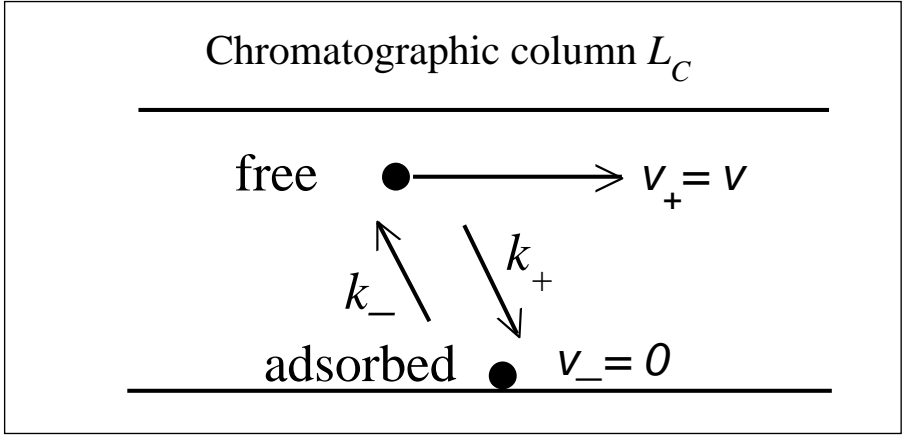}}
\caption{Schematic representation of a chromatographic column of length $L_c$.
Particles switch at random, with transition rate $k_+$, from the mobile state of velocity $v_+=v$
to an adsorbed, immobile state ($v_-=0$), and can return to the ``free" state
with a transition rate $k_-$. }
\label{figure3}
\end{figure}

{\em (a) Broadening of peaks in a chromatographic column}. \\
A chromatographic column is a very simple device for separating particles of different
characteristics; it consists of a cylindrical tube, where particles are drifted by a fluid, with mean velocities depending on their mobilities. So, at the end of the tube, one is expecting to receive neat ``delta-peaks" of identical particles. It was noted experimentally, however, that there is a broadening of these peaks, i.e., the identical particles do not arrive exactly at the same moment at the end of the column (despite all the precautions related to ensuring identical initial conditions at the beginning of the tube, etc.).
In the model proposed in
Refs.~\cite{giddings55} and \cite{giddings57} for this broadening phenomenon, for each species of
particles present in this sorting-device there are  random
switches, with a specific transition rate $k_+$, from a mobile state (in which particles are dragged with a specific velocity $v$ along the chromatographic column)
to an immobile, adsorbed state; they are afterwards randomly desorbed,
 with a characteristic desorbing rate $k_-$, see Fig.~\ref{figure3}
for a schematic representation.

This stochastic process can be described as
\begin{equation}
\dot{x}(t)=v/2+\xi(t)\,,
\end{equation}
where $x(t)$ designates the position of the particles along the column, and $\xi(t)$ is a DMN that takes the values $\pm v/2$ with transition rates $k_{\pm}$. This results in an asymptotic mean velocity of the particles
\begin{equation}
\langle \dot{x}\rangle =\frac{k_-}{k_++k_-}\;v\,
\end{equation}
and a dispersion of the particles around this drift profile (a ``broadening" of the profile) given by an effective diffusion coefficient
\begin{equation}
D_{\mbox{eff}}=\frac{k_+k_-}{(k_++k_-)^3}\; v^2\,.
\end{equation}
The separating efficiency of the chromatographic column is thus determined by its length $L_c$,  through the condition that  for each type of particles ``convection wins over dispersion",
\begin{equation}
L_c>\frac{2D_{\mbox{eff}}}{\langle \dot{x}\rangle}=\frac{2k_+v}{(k_++k_-)^2}\,,
\end{equation}
and also by requiring ``sufficiently different" asymptotic mean velocities of the
particles of different species. Note also that reaching the asymptotic regime
requires times $t \gg \tau_c=1/(k_++k_-)$, and therefore $L_c$ should
be actually much larger than the right-hand side of the above equation.\\

\begin{figure}[h!]
\quad \vspace{1.5cm} \\
\centerline{\psfig{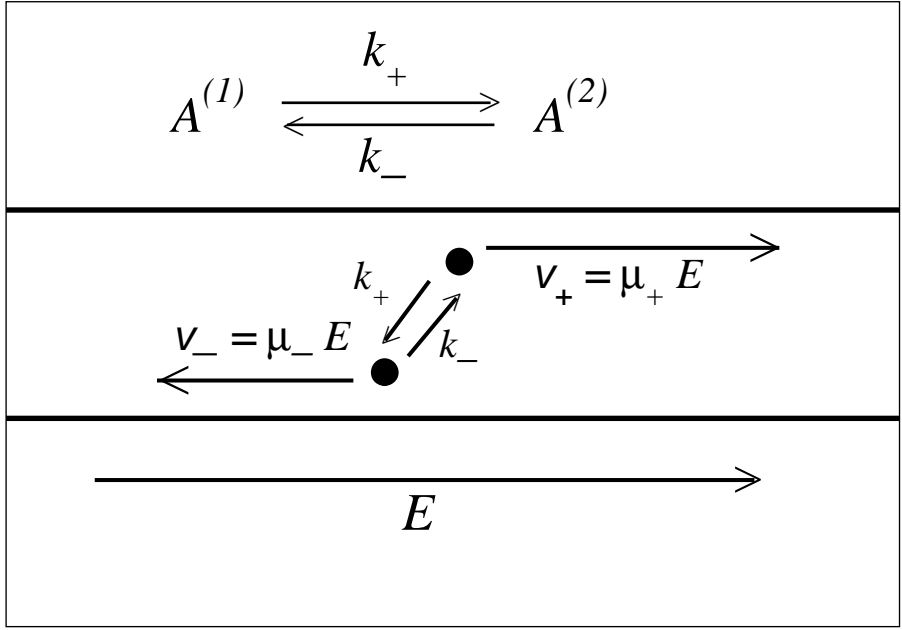}}
\caption{Schematic representation of an electrophoresis experiment. An ensemble of particles undergo random transitions between two states, $A^{(1)}$ and $A^{(2)}$,
that have different electrophoretic mobilities $\mu_{+}$,
respectively $\mu_-$.  By applying an external electric field
$E$ and by measuring the mean drift velocity of the particles, as well as the
diffusion around the mean drift position, one can obtain the values of the transition rates $k_{\pm}$
between the states $A^{(1,2)}$.}
\label{figure4}
\end{figure}

{\em (b) Measuring reaction rates through electrophoresis}. \\
In this system, as modeled in Ref.~\cite{mysels56}, the jumps (with  transition rates $k_{\pm}$)
between the two states of a DMN correspond
to a particle undergoing a chemical transformation between two configurations,
$A^{(1)}$ and $A^{(2)}$,  with different
electrophoretic mobilities $\mu_{\pm}$. See Fig.~\ref{figure4} for illustration.
Of course, by measuring  ``statically" the
concentrations $[A^{(1,2)}]$ of the two configurations, one can obtain the ratio of the transition rates  $k_{\pm}$, since at equilibrium $k_+ [A^{(1)}] = k_- [A^{(2)}]$.
When applying an external field,
one can measure ``dynamically" the average velocity of the electrophoretic peak,
\begin{equation}
\langle\dot{x}\rangle =
\displaystyle\frac{\mu_+k_-+\mu_-k_+}{k_++k_-}\;E
\end{equation}
(which contains the same information as the equilibrium measurements), but also
the dispersion of the particles around this peak, i.e., their effective diffusion coefficient
\begin{equation}
D_{\mbox{eff}}=\displaystyle\frac{k_+k_-(\mu_+-\mu_-)^2}{(k_++k_-)^3}\;E^2\,,
\end{equation}
and thus one can determine $k_+$ and $k_-$ separately.\\

\begin{figure}[h!]
\quad \vspace{1.5cm}\\
\centerline{\psfig{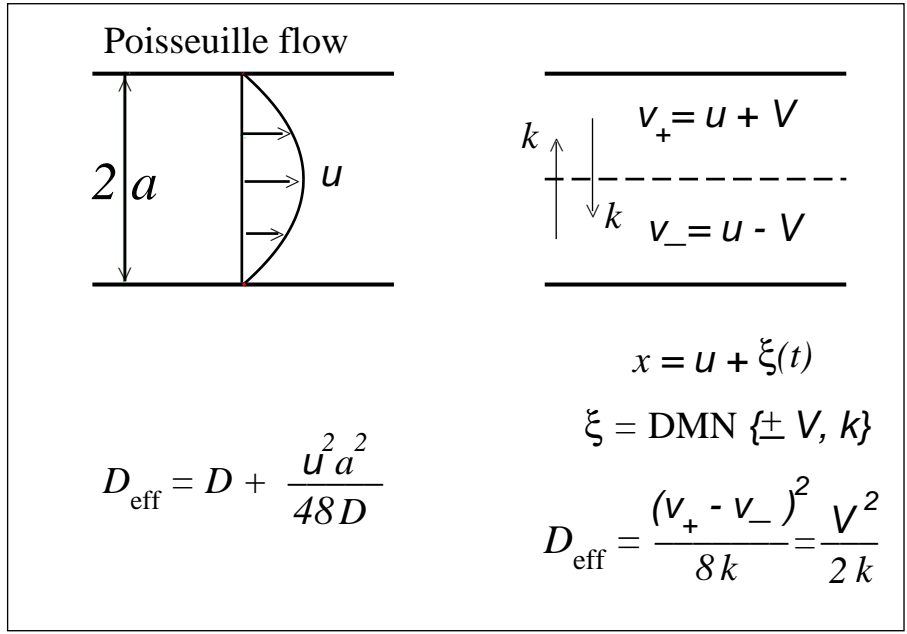}}
\caption{Schematic representation of a Poiseuille flow in a cylinder (left-hand side of the figure),
with the expression of the effective diffusion coefficient along the direction of the flow
as found theoretically by Taylor. The right-hand side of the figure represents
a caricature of the Taylor dispersion seen as a dichotomous diffusion, see the main text. Indicated are the
model stochastic process, and the resulting effective diffusion coefficient along the axis of the cylinder.}
\label{figure5}
\end{figure}

{\em (c) Taylor dispersion}. \\
In the fifties, Taylor investigated, both theoretically and experimentally, the motion of tracers in a Poiseuille flow in  cylindrical configuration,
see Refs.~\cite{taylor1,taylor2,taylor3}. Applications refer, e.g.,  to the motion of pollutants in rivers, etc.
In the long-time regime, he found that
the tracers are being dragged downstream along the $x$-axis of the cylinder, with the mean velocity $u$ of the flow. But besides that, the tracers are also dispersed around this drift peak, with an effective diffusion coefficient
$D_{\mbox{eff}}$ that is inversely proportional to their molecular diffusion
coefficient $D$. See the left-hand side of Fig.~\ref{figure5} for an illustration.
A very simple model of dichotomous diffusion introduced later, see Ref.~\cite{thacker75}, allows to capture the essential features of the Taylor dispersion. It consists of tracers that jump at random,
with transition rates $k$, between two layers of fluid with velocitis $u+V$
and $u-V$ respectively. This results in an effective diffusion coefficient
along the $x$-axis, $D_{\mbox{eff}}$ that is inversely proportional to the transition rate $k$.
One realizes that, as in the case of the Taylor dispersion, $D_{\mbox{eff}}$ is proportional to the typical time that is needed by a particle to sample all the available velocities
($a^2/D$ for the Taylor flow, and $k^{-1}$ for the dichotomous diffusion). Note also
the divergence of $D_{\mbox{eff}}$ when there is no transition between the layers of the fluid (i.e., when $D\rightarrow 0$ for Taylor dispersion, respectively $k\rightarrow 0$
for dichotomous diffusion); indeed, in this case particles with different velocities move apart from each other with a ballistic motion.\\

\begin{figure}[h!]
\quad \vspace{1.5cm} \\
\centerline{\psfig{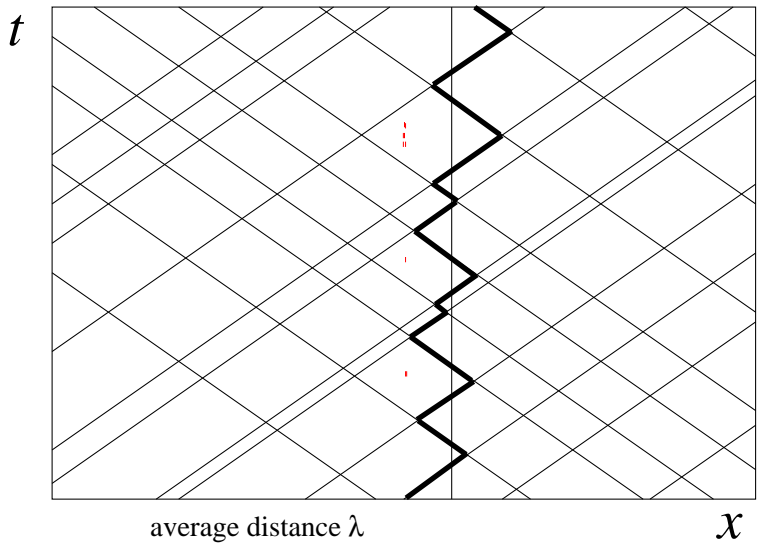}}
\caption{Schematic representation of one realization of the trajectories of the hard-point particles
of a Jepsen-McKean gas (see main text). The thick line represents the trajectory of a test-particle that undergoes a dichotomous diffusion due to the collisions with the other particles.}
\label{figure6}
\end{figure}

{\em (d) Jepsen-McKean gas}. \\
This is a very simple (although theoretically very productive)
model of a classical gas. It consists of elastic, hard-point particles that are initially randomly distributed  on a line
(with an average distance $\lambda$ between them),
and with randomly distributed initial velocities taking
two possible values,  $+ v$ and $-v$,
see Refs.~\cite{jepsen65,mckean67,protopopescu85}.
At collisions particles simply exchange trajectories, see Fig.~\ref{figure6}
for a schematic representation.
The velocity of a particle is thus  just $\dot{x}(t)=\xi(t)$, with the DMN
$\xi(t)$ taking the values $\pm v$ with transition rate $k=v/\lambda$.
For this system, the master equation for the probability densities $P_{\pm}(x,t)$
is nothing else but Boltzmann's equation (which is {\em exact} for this model!), and the probability distribution function
$P(x,t)$ (that obeys telegrapher's equation) is proportional to
the spatial density of particles.\\

\begin{figure}[h!]
\quad \vspace{1.5cm} \\
\centerline{\psfig{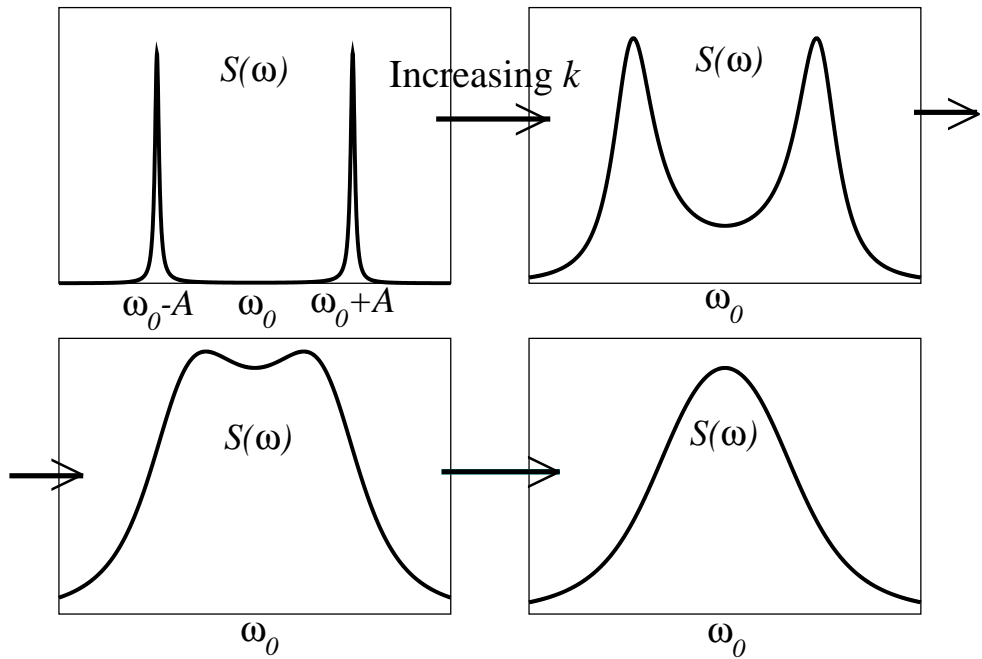}}
\caption{Schematic representation of the motional narrowing of the power spectrum of the random-frequency oscillator driven by  a DMN $\pm A$ (see the main text).
The spectrum, that is two-peaked for low transition rates $k$ of the DMN, becomes progressively single-peaked and narrower with increasing $k$. }
\label{figure7}
\end{figure}

{\em (e) Kubo-Anderson oscillator.}\\
 Finally, an example that has numerous
applications in spectroscopy (see, e.g., Ref.~\cite{mukamel} for a review),
is the so-called ``random-frequency oscillator" introduced in Refs.~\cite{anderson53,kubo54I,kubo54II,anderson54,kubo69}. It is  described by the stochastic evolution of a phase-like variable,
\begin{equation}
\dot{u}(t)=-i\,\left[ \omega_0 +\xi(t)\right]\,u\,,
\end{equation}
where, in one of the simplest variants, $\xi(t)$ is a DMN $\pm A$ with transition rate $k$.
The mean value of the phase variable
\begin{eqnarray}
\langle u(t)\rangle &=& \langle u(0) \rangle \langle \exp\left[-i \omega_0 t - i
\int_0 ^t \xi(t') dt' \right]\rangle \nonumber\\
&=&\langle u(0) \rangle e^{-i\omega_0 t-kt}\,
\left\{\cos(t\sqrt{A^2-k^2})+
\frac{k}{\sqrt{A^2-k^2}}\sin(t\sqrt{A^2-k^2})\right\},
\end{eqnarray}
leads to a power spectrum
\begin{equation}
S(\Delta \omega)=\frac{2{k}}{\pi}\;\frac{A^2}{[(\Delta
\omega)^2-A^2]^2+4{k^2} (\Delta \omega)^2}\,,
\end{equation}
with $\Delta \omega = \omega - \omega_0$. As illustrated in Fig.~\ref{figure7}, there
is a change in the shape of the spectrum with increasing frequency $k$
of the transition, the so-called  {\em motional narrowing}.
Such a random-frequency oscillator can describe, for example, the Larmor precession of a  spin in a random magnetic field;
or the thermal transition of a molecule between configurations with different
absorbtion frequencies, etc.
By measuring the motional narrowing in the power spectrum, one can obtain various information on the structure of the system under study. For example, from measurements on the NMR spectrum at different temperatures
(which means for different Arrhenius-like transition rates between different structural configurations),  one can determine the energy barriers
between the various configurations of a molecule -- see, e.g., two early-day papers
of J. A. Pople, Refs.~\cite{pople1} and \cite{pople2}.

\subsubsection{Dichotomous diffusion and Quantum mechanics}

Let us now mention briefly two types of connections with quantum mechanics problems.

{\em (a) Relativistic extension of the analogy between quantum mechanics and \\
Brownian motion.}

As  first remarked in Ref.~\cite{gaveau84}, the telegrapher's equation~(\ref{telegrapher}) can be converted into the Klein-Gordon equation
through a simple transformation,  that suggests a formal connection with relativistic wave equations. More precisely, setting:
\begin{equation}
P_{\pm}(x,t)=e^{-kt} \,\psi_{1,2}(x,t)
\end{equation}
in Eq.~(\ref{telegrapher}), one finds that each of the ``components" $\psi_{1,2}$ obeys
the Klein-Gordon equation:
\begin{equation}
(\partial_t^2-A^2\partial_x^2-k^2)\psi_{1,2}=0\,.
\label{coupled1}
\end{equation}
Or, by substituting in the master Eq.~(\ref{ppm}), one finds the coupled equations for the components $\psi_{1,2}$:
\begin{equation}
(\partial_t\pm A\partial_x)\psi_{1,2}=k\psi_{2,1}\,.
\end{equation}
On the other hand, the Dirac equation for a free relativistic particle of rest mass $m_0$
reads
\begin{equation}
(i\gamma^{\mu}\partial_{\mu}-m_0c/\hbar)\psi=0
\end{equation}
($c$ is the speed of light, and $\hbar$ the reduced Plancks's constant).
In $(1+1)$ dimensions $\partial_{\mu}=(c^{-1}\partial_t,\partial_x)$ and
using Weyl's representation
\begin{equation}
\gamma^0=\left(\begin{array}{cc} 0 & 1 \\ 1 & 0 \end{array} \right)\,,\quad
\gamma^1=\left(\begin{array}{cc} 0 & -1 \\ 1 & 0 \end{array} \right)\,,
\end{equation}
the Dirac's equation for $\psi=(\psi_1,\,\psi_2)$ reduces to
\begin{equation}
(\partial_t\pm c\partial_x)\psi_{1,2}=(m_0c^2/i\hbar)\psi_{2,1}\,.
\label{coupled2}
\end{equation}
Of course, these are exactly Eqs.~(\ref{coupled1}), provided that we make the following identifications:
\begin{equation}
A \leftrightarrow c\,,\quad k \leftrightarrow m_0c^2/i\hbar\,.
\end{equation}
The factor $i$ that appears above reflects the familiar analytical continuation to ``imaginary time" that appears in the usual passage from the diffusion equation to the Schr\"odinger equation (or from the diffusion kernel to the quantum mechanics propagator). So, in this picture, a Dirac particle in one dimension, with helicity amplitudes $\psi_{1,2}$ moves back and forth on a line, at the speed of light (Zitterbewegung!),
and at random times it flips both the direction of propagation and the chirality.

This analogy was generalized later in Ref.~\cite{balakrishnan05} to the case of
a velocity-biased dichotomous diffusion.  The transition rates
between the $+A$ state and the $-A$ state are unequal, $k_+\neq k_-$, thus resulting in a net drift velocity, see Ref.~\cite{filliger04} for a detailed discussion of this process.
The modifications in the master equation and its solutions can be interpreted in terms of a Lorentz transformation to a frame moving precisely with this drift velocity.
As a consequence, in the connection to the Dirac equation this results in a modification of the correspondence between the rest mass of the Dirac particle and the frequency
of direction reversal in the dichotomous diffusion, the latter being corrected
by the time-dilatation factor corresponding to the drift velocity.

The analogy between stochastic differential equations and quantum mechanical waves has been developed further in the context of stochastic quantum mechanics (see Ref.~\cite{balakrishnan05} for some pertinent references). However, many nontrivial open questions still remain (e.g., the extension of the analogy between diffusive processes and relativistic wave equations to more than one spatial dimension, to cite only an example). \\

{\em (b) Quantum random walks and dichotomous diffusion.}

In Ref.~\cite{blanchard04}  one considers a one-dimensional
space and time-discrete quantum random walker
driven by a Hadamard ``coin-tossing process".
Namely, at each time step the two-component wave function $\psi(n,t)=(\psi_1(n,t),\,\psi_2(n,t))$
of the particle on a discrete lattice 
at position $n$ suffers a modification of the chirality
and a jump to the neighbour sites,
according to the rules:
\begin{eqnarray}
&&\psi_1(n,t+1)=-\frac{1}{\sqrt{2}}\psi_1(n+1,t)+\frac{1}{\sqrt{2}}
\psi_2(n-1,t)\,,\nonumber\\
&&\psi_2(n,t+1)=\frac{1}{\sqrt{2}}\psi_1(n+1,t)+\frac{1}{\sqrt{2}}
\psi_2(n-1,t)\,.
\end{eqnarray}
In the continuum space and time limit, the probability distribution for the quantum particle (i.e., the square of the wave function) is shown to obey a hyperbolic equation that is similar to telegrapher's equation obtained for a dichotomous diffusion with space-dependent transition rates between the two states of the
driving DMN. This leads to an asymptotic ballistic behavior of the mean square ``displacement" of the quantum particle,
contrary to its classical counterpart
(an effect of quantum interferences). The generalization to a larger class of unitary transformations of the chirality than the Hadamard coin-tossing will lead, in the continuum limit, to a larger class of hyperbolic equations describing piecewise deterministic motions.
Solving these equations is, however, a very difficult task~\cite{blanchard04}.\\

In view of the already-discussed difficulty and thus scarcity of exact results for time-dependent problems, but mainly in view of the relevance for practical purposes,
we shall turn below to the  asymptotic, stationary regime of the dichotomous flows.

\section{DICHOTOMOUS FLOWS: STATIONARY SOLUTIONS WITHOUT UNSTABLE CRITICAL POINTS}

The first subsection will be devoted to the definition and general properties of simple
stationary dichotomous flows on the real axis, in the absence of unstable critical points.
We discuss afterwards
the appearance of noise-induced transitions on the particular example of the genetic model.
A comparison is made between the effects of the DMN and those of a deterministic periodic
dichotomous perturbation.
In the third paragraph we are considering a simple ensemble of coupled
dichotomous flows, and analyse the mechanism behind the onset of a global
instability in this model of a spatially-extended system. The DMN and the
deterministic periodic forcing are contrasted again in this context.
Finally, in the fourth
subsection we are ready to address the problem of the nonequilibrium
phase transitions induced
by DMN in a mean-field type of model. Comparison with the phase transitions
induced by GWN allows to emphasize  specific features
(multistability, hysteretic behavior, etc.) related to the colour
of the DMN noise.

\subsection{Generalities on stationary dichotomous flows}

The vaste majority of the studies on dichotomous flows refer to the {\em stationary},
long-time behavior.  We shall briefly present here the most important,
generic results for flows $f_{\pm}(x)$ that are defined and ``sufficiently smooth"
(at least continuous) on the whole real axis,
without any specific  periodicity properties, and {\em without  unstable critical
points}. This last condition, that avoids important mathematical difficulties 
(that will be revisited in Sec.~V) was the only one considered in the ``classical"
literature on the subject. We will admit it throughout this Section.

\begin{figure}[h!]
\quad \vspace{1.5cm} \\
\centerline{\psfig{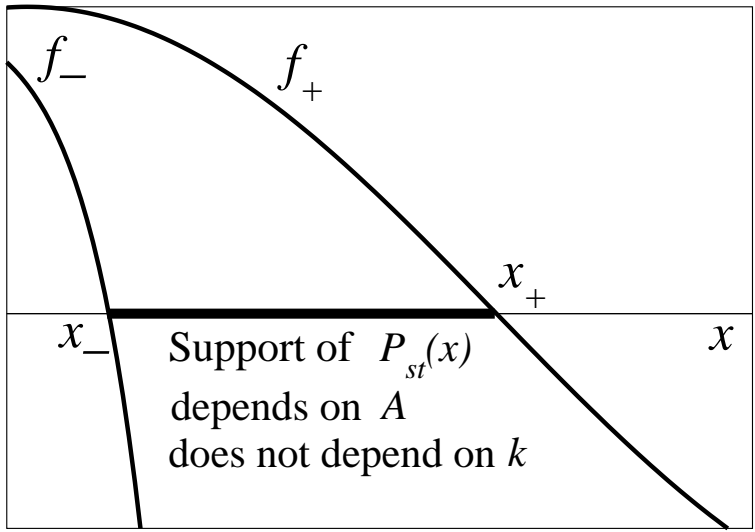}}
\caption{The attractors $x_-$ and $x_+$ of the alternate flows $f_-(x)$, respectively $f_+(x)$ define the compact support of the stationary probability distribution $P_{st}(x)$ of the dichotomous flow. See the main text.}
\label{figure8}
\end{figure}

Note, however, that each of $f_{\pm}(x)$ may have, eventually,  one
{\em stable} critical point (of course,  in view of the supposed continuity and absence of unstable critical points of $f_{\pm}(x)$, no more than one fixed point is allowed for each $f_{\pm}(x)$).
The only non-trivial situation that can therefore be considered in this context appears when each of the alternate ``$+$" and ``--" flows does have a {\em stable}
critical point, let us call them $x_+$, respectively $x_-$~\footnote{There are two other
possible situations, but they are both trivial:
(i) When one single flow has a stable critical point, the asymptotic distribution reduces to a $\delta$-peak at this point. (ii) When none of the two flows has fixed points, the initial distribution reduces progressively to zero throughout the real axis -- the particles escape to infinity. }:
\begin{equation}
f_{\pm}(x_{\pm})=0\,, \quad \mbox{with} \quad f'_{\pm}(x_{\pm})<0
\end{equation}
(the prime denotes the derivative with respect to $x$).
 Due to the competition between these attractors, the asymptotic motion of the particles settles down in an alternate flow between the two attractors, a process called
{\em dynamic stability}, which leads to a nontrivial  stationary probability distribution $P_{st}(x)$. Thus the attractors  define the compact support of  $P_{st}(x)$, see Fig.~\ref{figure8}.
The support depends on the amplitude of the DMN, however it is clear that it does not depend on the transition rates. Indeed, due to the exponential distribution of the
waiting times in each  state of DMN, the particle can persist into
one of the flows till reaching the corresponding attractor, whatever
its initial position with respect to this attractor.

The stationary master equation for $P_{st}(x)$  can be easily solved, see e.g.
Ref.~\cite{horsthemke84}, and leads to the following expression
(for $x_- \leqslant x\leqslant x_+$):
\begin{eqnarray}
P_{st}(x)d x&=&{\cal N} \left(\frac{d x}{|f_+(x)|}+\frac{d x}{|f_-(x)|}\right)\times\nonumber\\
&& \times \left\{k\exp[-k\,T_+(x_-,x)]\right\}\;
\left\{k\exp[-k\,T_-(x_+,x)]\right\}\,,
\end{eqnarray}
where
\begin{equation}
T_{\pm}(u,x)=\int_u^x\frac{dz}{f_{\pm}(z)}\,,
\end{equation}
and ${\cal N}$ is a normalization factor.
The structure of this expression has a simple intuitive interpretation: \\
(a) Particles that are found in the interval $dx$ around $x$ can be either in the $+$, or in the -- flow, spending in $dx$ a time that is inversely proportional
to their velocity. From here the term $\left({d x}/{|f_+(x)|}+{d x}/{|f_-(x)|}\right)$.\\
(b) In order to reach the point $x$, particles: (i) either  come from $x_-$ and are driven by the $f_+(x)$ flow (towards its attractor); this takes a time $T_+(x_-,x)$, and the probability for the DMN to persist in the $+$ state during this time is
$\left\{k\exp[-k\,T_+(x_-,x)]\right\}$; (ii) or come from $x_+$ and
are driven by the $f_-(x)$ flow; this takes a time $T_-(x_+,x)$
and happens with probability $\left\{k\exp[-k\,T_-(x_+,x)]\right\}$.

Near the attractors, $P_{st}(x)$ has a typical behavior
\begin{equation}
P_{st}(x) \sim \left|x-x_{\pm}\right|^{-1-k/f'_{\pm}(x_{\pm})}\,,
\end{equation}
which reflects the competition between two time scales, namely that of the switching
between the two flows  and that of the dynamics in the vicinity of the attractors. More precisely: (a)  If the transition rate $k$ is low  (${k}/{|f'_{\pm}(x_{\pm})|}<1$), there is
an accumulation of particles at the stable fixed point (and this results in an integrable divergence of the probability distribution function); (b) If the transition rate $k$ is large,
the alternance of dynamics is sufficiently efficient in sweeping-out the effect of the attractor.  $P_{st}$ becomes zero at the borders, either with a divergent slope
(for $1<{k}/{|f'_{\pm}(x_{\pm})|}<2$) or with a zero slope for
${k}/{|f'_{\pm}(x_{\pm})|}>2$.\\

Another point of interest is the location of the {\em maxima} $x_m$ of $P_{st}(x)$, and their dependence on the parameters of the noise. Why are these maxima important? Because (in view of the ergodicity of the system) they offer indications on the values of the stochastic variables that are most-likely encountered in a finite-time
measurement on the system. The equation giving the extrema
\begin{equation}
f(x_m) - Dg(x_m) g'(x_m) + \tau_c \; \left[ 2 \, f(x_m) \, f'(x_m) -
\, f^2(x_m) \, \frac{g'(x_m)}{g(x_m)} \right]=0
\end{equation}
has several contributions: the first term in the left-hand side  corresponds to the deterministic
(noiseless) steady-state. The second term is related
to the multiplicative nature of the DMN ($g'(x)\neq 0$), and persists in the
limit of a Gaussian white noise (where it is known as the ``Stratonovich spurious drift",
see, e.g, Ref.~\cite{vandenbroeck97}). Finally, the last term on the left-hand side
is due to the finite correlation time of the DMN~\footnote{ In relation with the already-mentioned differences in their statistical properties, note that this term is different from the
term corresponding to an Ornstein-Uhlenbeck process.}, and, of course, is absent
in the  GWN-limit.\\

The nonlinearity of $f_{\pm}(x)$ leads to the possibility of multiple maxima of $P_{st}(x)$.  Moreover, by modifying the parameters of the noise, one can vary the
position of these maxima, or can even create new maxima
(that are absent in the absence of the noise).
The macroscopic state of the system can thus undergo a
{\em qualitative} change under the influence of the noise.
Coloured-noise processes can give rise to states that disappear in
the limit of a white noise. These very simple
remarks are at the basis of the celebrated {\em noise-induced transitions},
see Ref.~\cite{horsthemke84} for a review.  Note, however, that this change in the shape of $P_{st}(x)$ {\em does not lead to a breaking of ergodicity}
(roughly, the initial state of the stochastic system does not influence on its asymptotic behavior).  Also, sometimes, these noise-induced transitions reduce mathematically to a nonlinear change of the stochastic variable.

\subsection{The genetic model: a case-study of transition induced by DMN}

One of the best candidates to illustrate the concept of noise-induced transition is, without doubt, represented by the so-called {\em genetic model}, see Refs.~\cite{arnold78} and \cite{horsthemke84}.
Indeed, in the absence of the noise this model  exhibits a unique
steady state that is stable: the system is not capable of any transitions under deterministic conditions.
We shall show that, depending on the parameters of the noise, new macroscopically observable states (in the sense  described in the paragraph above)
can be generated~-- ~i.e., {\em purely} noise-induced transitions
can take place in the system.

The equation of  the model is
\begin{equation}
\dot{x}=(1/2 - x) +\lambda x(1-x)\,,
\label{gen}
\end{equation}
where $x(t)$ is the state variable (that takes values between $0$ and $1$),
and $\lambda$ is the control parameter which models the coupling of the
system to its environment.

This very simple model was initially
introduced on a purely theoretical basis, see Ref.~\cite{arnold78}.
A  realization of it is given by the mean-field description of
the coupled auto-catalytic reactions:
\begin{equation}
A+X+Y
\left.
\begin{array}{c}
k_1\\
\rightleftarrows \\
k_2
\end{array}
\right.
2Y+C
\end{equation}

\begin{equation}
B+X+Y
\left.
\begin{array}{c}
k_3 \\
\rightleftarrows \\
k_4
\end{array}
\right.
2X+D
\end{equation}
The reactions conserve the total number of $X$ and $Y$ particles:
\begin{equation}
X(t)+Y(t)=N=\mbox{constant}
\end{equation}
 Considering
\begin{equation}
 \alpha=\frac{k_2C}{k_2C+k_4D}=1/2
 \end{equation}
 and
 \begin{equation}
 \lambda=\frac{k_3B+k_4D-k_1A-k_2C}{k_2C+k_4D}\,,
 \end{equation}
 one obtains the genetic equation~(\ref{gen}) for the variable $x=X/N$. The
 concentrations $A$, $B$,  $C$,  and $D$ are externally-controlled and determine the value
 of $\lambda$.

Later on it was shown
that this model can be used to describe  quite realistically a mechanism of
genetic evolution in a two-genotype population  -- and this is where the name of the model came from  (see Ref.~\cite{horsthemke84} for detailed comments).
In this context, $x$ represents the frequency of one of the two genotypes.
The term $(1/2-x)$ corresponds to a mutation between genotypes,
that tends to equalize the frequency of their expression.
Finally,  the term $\lambda x(1-x)$
corresponds to a natural selection mechanism that tends
to favor the genotype that is ``best adapted" to the environment;
$\lambda$ represents the ``selection rate", that is environment-dependent and
that can be either positive or negative.
The deterministic ($\lambda$ = constant) steady-state is just
\begin{equation}
x_{\mbox{det}}=[\lambda-1+(\lambda^2+1)^{1/2}]/(2\lambda)
\end{equation}
and, as it was already mentioned, one can easily verify that it is asymptotically, globally stable.

\begin{figure}[h!]
\quad \vspace{1.5cm} \\
\centerline{\psfig{file=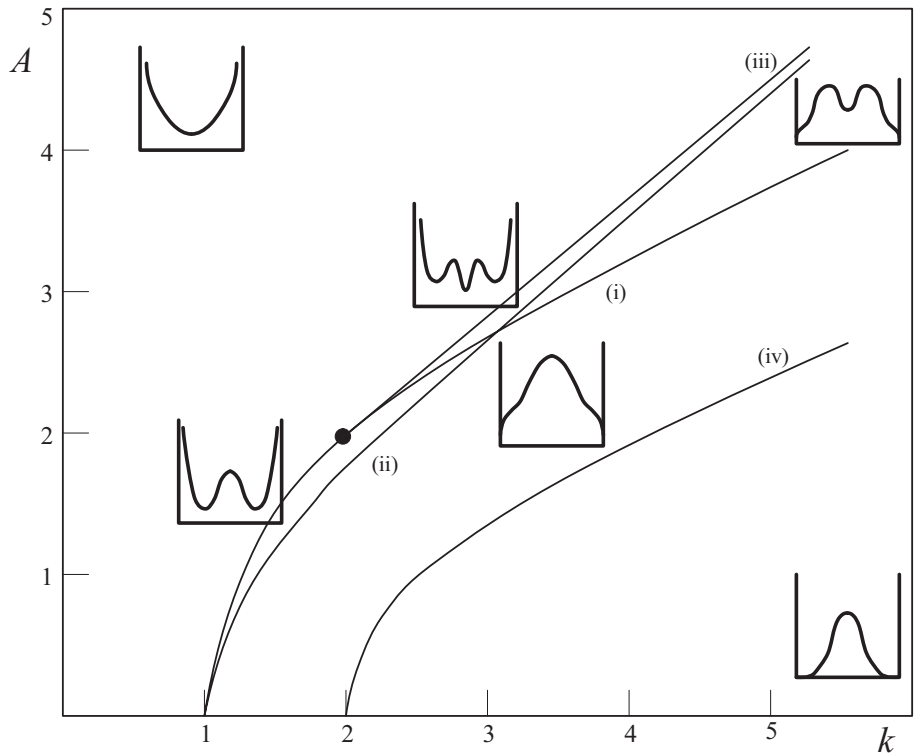,width=12cm}} \vspace*{8pt}
\caption{The phase diagram of the genetic model driven by a symmetric DMN in the plane of the parameters $(k, \,A)$ of the DMN. The borderlines correspond, respectively, to:
(i) $A=2\sqrt{k-1}$ for $k\geqslant 1$ ; (ii) $A=\sqrt{k^2-1}$ for $k\geqslant 1$; (iii) $A=k$ for $k\geqslant 2$; and
(iv) $A=\sqrt{k^2/4-1}$ for $k\geqslant 2$. The  shape of $P_{st}(x)$ for the
different regions is represented qualitatively, illustrating the peak-splitting and the peak-damping effects, see the main text.}
\label{figure9}
\end{figure}

Let us now try to model  the effects of a fluctuating environment
on the genotype dynamics.
This can be done by assigning a fluctuating
  ``selection rate" $\lambda$;   for example $\lambda=\xi(t)$ can be a symmetric DMN
$\{\pm A,\, k\}$ (such that on the average no genotype is favored by the natural selection),
see Ref.~\cite{horsthemke84}.
The corresponding stochastic differential equation
\begin{equation}
\dot{x}=(1/2-x) +\xi(t) \,x(1-x)
\label{genstoc}
\end{equation}
leads to a stationary probability distribution function $P_{st}(x)$ that can be studied
in full analytical detail. Putting together the information on the support of $P_{st}(x)$ (the region between the attractors of the alternate ``+" and ``--" dynamics);
on its maxima; on its behavior (and that of its derivative, see above) near the borders, one can obtain  the ``phase diagram" of the system in the plane of the parameters $k$ and $A$ of the DMN, see Fig.~\ref{figure9}.

The simplicity of the stochastic structure of the DMN (that was invoked in the Introduction)
allows us to decipher the two mechanisms at  work in fashioning the shape of the stationary probability distribution $P_{st}(x)$. The first one is a {\em peak-damping mechanism}, which is a generic disordering effect of the noisy environment. The second
mechanism is a {\em peak-splitting mechanism}, signature of the two-valued DMN,
and source of the noise-induced qualitative changes in the shape of $P_{st}(x)$ (i.e., of the noise-induced transitions). Recall, once more, that there is no ergodicity breaking in the
system. Note, moreover, {\em the disappearance of the signatures of a
stochastic behavior for too large or too low transition rates of the noise,
 which is a generic effect
for all dichotomous flows}. This is  very simple to explain  intuitively:\\
(a) In the first case (``large" $k$-s) the system ``does not have time"
to adapt to the rapidly-changing instantaneous value of the noise, 
and  the stochastic effects are simply smeared out.
(From here results the single deterministic peak of $P_{st}(x)$ 
in Fig.~\ref{figure9});\\
(b) In the latter case (``small" $k$-s) half of the systems of the statistical
ensemble are evolving along $f_{+}(x)$, and half along $f_{+}(x)$, 
practically without interchange between these subensembles.
(From  here result the two deterministic peaks of 
$P_{st}(x)$ in Fig.~\ref{figure9}.)\\

\subsubsection{What is the difference with a dichotomous periodic forcing?}

\begin{figure}[h!]
\quad \vspace{1.5cm} \\
\centerline{\psfig{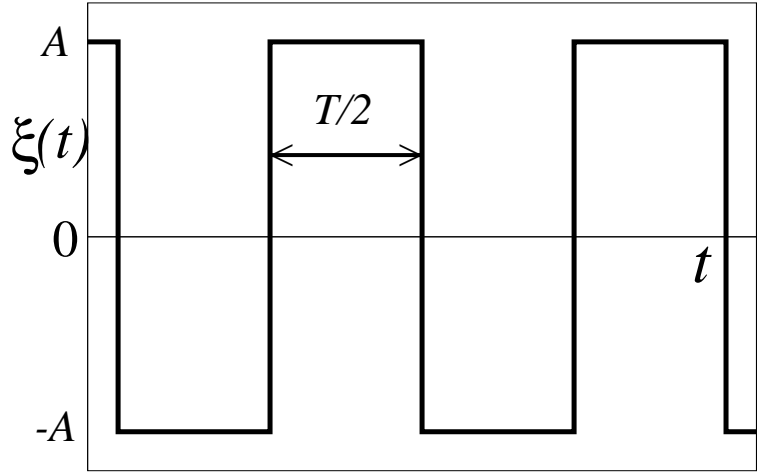}} \vspace*{8pt}
\caption{A deterministic periodic dichotomous forcing $\xi(t)$ of amplitude $A$ and  period $T$.}
\label{figure10}
\end{figure}

\begin{figure}[h!]
\quad \vspace{1.5cm} \\
\centerline{\psfig{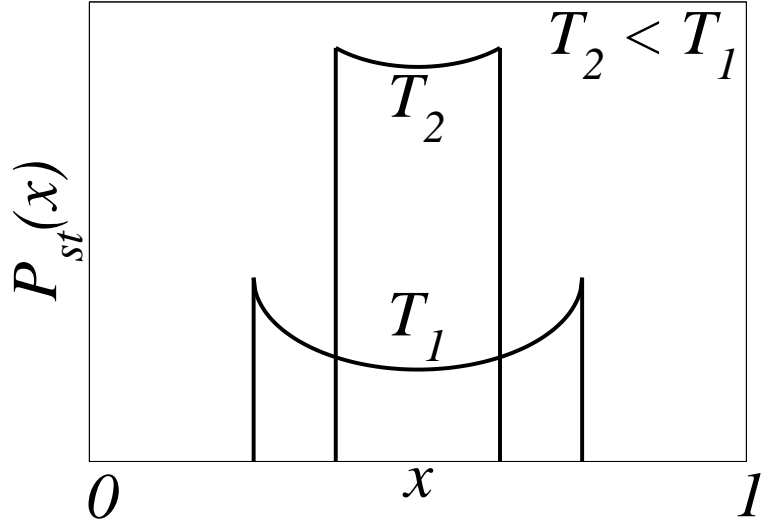}} \vspace*{8pt}
\caption{Qualitative profile of the stationary probability distribution $P_{st}(x)$
for the case of a deterministic periodic dichotomous forcing $\xi(t)$ in the
genetic model. We considered two perturbations of the same amplitude $A$
and different periods $T_1$ and $T_2$. $P_{st}(x)$ is always bimodal and its support depends on both the amplitude and the period of the perturbation.}
\label{figure11}
\end{figure}

In this context, people also addressed the following question: what is the
difference (if any) in the transitions induced by random (DMN) and periodic dichotomous
perturbations? Let us consider  this problem in the context of the genetic model,
following Ref.~\cite{doering85}, and  take in Eq.~(\ref{genstoc})
$\xi(t)$ to be a dichotomous periodic forcing as shown in Fig.~\ref{figure10}.
The stationary probability distribution $P_{st}(x)$
(corresponding to random initial conditions in the variable
of the system and in the phase of the perturbation)
is given (with obvious notations) by:
\begin{equation}
P_{st}(x)dx=\frac{1}{T}(dt_++dt_-)=
\frac{1}{T}\left(\frac{dx}{|f_+(x)|}+\frac{dx}{|f_-(x)|}\right)\,.
\end{equation}
One notices that its  support  depends on both the amplitude $A$ and, contrarily to the case of the DMN, also on the period $T$ of the perturbation (which limits the time the system spends in the ``+", respectively ``--" states). But,  more important, $P_{st}(x)$ is
{\em always bimodal and qualitatively independent of the parameters
of the perturbation}, see Fig.~\ref{figure11}. This means that {\em no transitions are induced by the periodic perturbation}.
We can infer from this, as done in detail in Ref.~\cite{doering85},
the importance of the {\em exponentially rare,
``large excursions" of the DMN} (i.e., long intervals of time when it persists
in keeping the same value); these are reflected in the low-frequency part of the
power spectrum of the DMN perturbation. This part is missing in the spectral
distribution of the deterministic periodic forcing (which is, actually, a succession of $\delta$-peaks), and in some circumstances (like in the genetic model) it
can be important
in determining the overall behavior.

One is thus tempted to extrapolate and state that it looks like the behavior induced by the DMN is ``richer" than that induced by a periodic perturbation. We shall question this point further with the next case-study, by introducing an example of {\em coupled dichotomous flows}. This will also allow us to make a first step
towards the concept of
{\em noise-induced nonequilibrium phase transition} in spatially-extended systems.

 \subsection{Overdamped parametric oscillators: a simple case-study for perturbation-induced instabilities}

Over the past two-three decades there has been an increasing interest in the nonequilibrium behavior and the role of noise on {\em spatially extended systems} modeled as {\em ensembles of simple dynamical units coupled to each other}. Noise was longly-thought to be only a source of disorder  -- as suggested by many day-to-day situations, as well as by equilibrium statistical mechanics. However, it is now acquired  wisdom  that in out-of-equilibrium, nonlinear systems noise can actually lead to the {\em creation of new, ordered states}, i.e., to {\em noise-induced phase transitions}.
These represent a result of the {\em collective, noisy, and generically nonlinear evolution} of the component units, and are absent in the absence of the noise.
Moreover,  the  collective behavior is usually {\em qualitatively} different from that of the single dynamical units.

The first step in the onset of a phase transition is the
instability of the reference state, and the collective effects are very important at this
stage, either by promoting the instability, or by preventing/delaying it.
 We shall illustrate the onset of the instability on the example of
 a system consisting of coupled  {\em parametric oscillators}, which are widely-spread  simple dynamical units~\footnote{Indeed, as discussed in
 detail in Refs.~\cite{bena99,vandenbroeck00,bena02}, many linear and nonlinear, deterministic and stochastic systems that exhibit {\em energetic instabilities} can be modeled as parametric oscillators which undergo the parametric resonance phenomenon.}. For simplicity, we shall
 consider only the inertialess case (and we refer the reader to Refs.~\cite{bena99,vandenbroeck00,bena02} for the case with inertia, and for further details and discussions).

Once this step overtaken, we can proceed in Sec.~IV.D with a more ``standard" discussion of a {\em phase transition induced by} DMN in a system of coupled dynamical flows. We shall moreover address specific effects related to the colour of the noise
as compared to the GWN case.

 \subsubsection{A single oscillator}
 Let us start by considering first the dynamics of a single unit,
 namely a single scalar variable decaying
 at a rate that is parametrically modulated, see Ref.~\cite{vandenbroeck98}:
 \begin{equation}
 \dot{x}(t)=\left[-1+\xi(t)\right]x(t)\,.
 \label{malt}
 \end{equation}
In the spirit of the comparison presented
in Sec.~IV.B.1, we shall consider $\xi(t)$ to be either
a symmetric DMN $\{\pm A,\,k\}$,  or a deterministic periodic perturbation (of given amplitude and period) with a randomly uniformly distributed initial phase.

\begin{figure}[h!]
\quad \vspace{1.5cm} \\
\centerline{\psfig{file=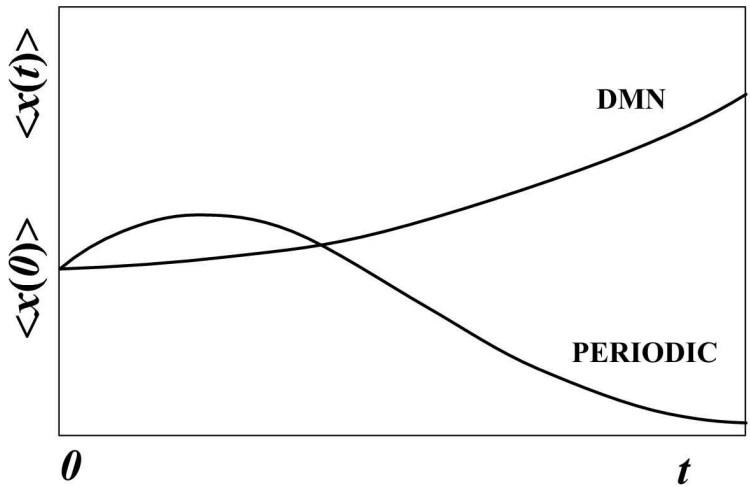,width=10cm}} \vspace*{8pt}
\caption{Qualitative features of the temporal
evolution of the first moment $\langle x(t)\rangle$ of the model
considered in Eq.~(\ref{malt}), for both a DMN and a deterministic
periodic perturbation. Note in the case of the periodic perturbation
the transient increase of $\langle x(t)\rangle$ above its initial
value.} \label{figure12}
\end{figure}

One can obtain analytically the temporal behavior of the first moment $\langle x(t)\rangle$
(where $\langle ...\rangle$ designates the mean over the realizations of the DMN in the first case, and over the initial phase in the case of the deterministic perturbation), with the typical profiles illustrated qualitatively in Fig.~\ref{figure12}.

Whatever the parameters of the deterministic periodic perturbation~\footnote{Except for the case of the quenched disorder $T\rightarrow \infty$ with $A>1$.}, {\em after a transient increase above the initial value},  the mean tends asymptotically to zero: the absorbing state $x=0$ is a global attractor.

For the case of the DMN, if the amplitude $A >\sqrt{2k+1}$ then the mean diverges in time~\footnote{Of course, the mean goes asymptotically to zero for $A <\sqrt{2k+1}$.}, although the absorbing state is still a global attractor, i.e.,
the probability distribution $P(x,t)$ tends asymptotically to $P_{st}(x)=\delta(x)$. The reason for this
counter-intuitive behavior has to be looked for into the existence of the extremely rare realizations of the DMN that lead to extremely large excursions
of $x$ away from zero. Even  if the latter ones are exponentially few, their weight
in the calculation of the mean is exponentially large, and can thus finally lead to the reported divergence of the mean. Although it cannot be argued that this represents
a noise-induced transition in the sense discussed in Secs.~IV.A and IV.B, it is
nevertheless an essential qualitative change in the transient stochastic behavior of the system.

\subsubsection{Coupled parametric oscillators}

\begin{figure}[h!]
\quad \vspace{1.5cm} \\
\centerline{\psfig{file=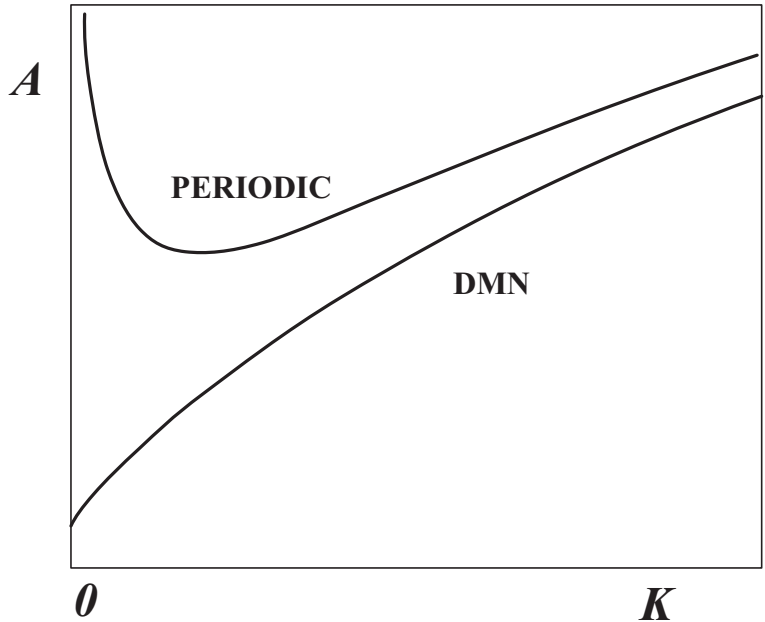,width=10cm}} \vspace*{8pt}
\caption{The instability borders in the $(K,\,A)$ plane  for an ensemble of globally, harmonically coupled overdamped parametric oscillators,  with a DMN, respectively a deterministic periodic parametric perturbation (qualitative representation;
see the main text). Note that the instability is re-entrant with respect to the coupling strength $K$.}
\label{figure13}
\end{figure}

Consider now $N$ such identical overdamped oscillators that are coupled in the simplest
possible way, namely an {\em all-to-all harmonic coupling}:
\begin{equation}
\dot{x}_i=[-1+{\xi_i(t)}]x_i-\sum_j K_{ij}(x_i-x_j)\,,
\end{equation}
with $i=1,...,N$ and $K_{ij}=K/N$. Let us take the thermodynamic limit $N\rightarrow \infty$ and  write down the evolution equation for
a generic oscillator (dropping the subscript $i$ for simplicity of notation)~\footnote{We admitted 
here, as a result of the law of large numbers, that
$\langle x(t)\rangle = \displaystyle\sum_i x_i(t)/N$ is a self-averaging macroscopic intensive variable, and that its value  coincides with the ensemble average over the realizations of $\xi(t)$.}:
\begin{equation}
\dot{x}(t)=[-1+{\xi(t)}]x-K[x-\langle x(t)\rangle]\,.
\end{equation}
One can then easily deduce the {\em instability border} that separates the asymptotic absorption regime $\langle x(t)\rangle \rightarrow 0$ from the
asymptotic explosive regime $|\langle x(t)\rangle |\rightarrow \infty$. This is  illustrated on Fig.~\ref{figure13}
in the plane of the parameters $A$ (amplitude of the perturbation)
and $K$ (coupling constant), for a fixed value of the transition rate $k$ for the DMN, respectively of the
period $T=1/(2k)$ for the deterministic forcing.

For the DMN, this border is simply given by:
\begin{equation}
A=\sqrt{1+2k +K}\;;
\end{equation}
therefore, the origin of the global instability of $\langle x(t)\rangle$
is very clear  in this case: if each oscillator is unstable, the mean will also be unstable. However, note that {\em coupling has a stabilizing effect in this context}, which means that you need a larger
amplitude $A$ of the noise in order to get unstable the center-of-mass of the oscillators.

For the periodic modulation, the instability border is given by
the following implicit equation:
\begin{eqnarray}
\frac{8A^2k K}{[(1+K)^2-A^2]^2}&&\left[\cosh\left(\frac{A}{4k}\right)-
\cosh\left(\frac{1+K}{4k}\right)\right]
\nonumber\\
&&=
\left[1-\frac{K(1+K)}{(1+K)^2-A^2}\right]\sinh\left(\frac{1+K}{4k}\right)\,.
\end{eqnarray}
So, in this example, things look to be more spectacular in the case of the
deterministic periodic forcing (as compared to the DMN case).
 Indeed, although {\em almost} each individual oscillator is asymptotically stable (i.e., $x(t) \rightarrow 0$ with probability almost 1), we have, however, a diverging mean. The reason is the existence of the initial transient increase of the mean (see the previous paragraph). This is due to the presence of a pool (that is permanently changing) of ``exceptional" individuals that make very large excursions away from zero. If the coupling is not too weak, and not too strong either, then it is promoting this initial transient increase, and these ``exceptional" individuals are thus leading the mean. So, in this situation, coupling may have a destabilizing effect! As expected, however,
the transition is {\em re-entrant}~\footnote{Re-entrance (i.e., the fact that the transition appears only for certain range of parameters of the noise or of the deterministic part of system's dynamics) is generic in out-of-equilibrium phase transitions. However, it is rather rare in equilibrium situations (examples include some $XY$ helimagnets and  insulator-metal transitions).} with respect to the coupling strength $K$
(a too large coupling do not allow for large excursions; while a too weak coupling does not allow the exceptional individuals to pass their role to the others, i.e., the permanent refreshing of the leading pool).

This simple example allowed us to make an important step in understanding the origin
of the onset of a global instability that may lead to a phase transition in an extended system: it is an initial, transient, local  instability in the system that is promoted and is ``not let to die out" by the coupling between the dynamical units of the system.
As discussed in a very  pedagogical way in Ref.~\cite{vandenbroeck97}, this is
a generic mechanism behind the onset of nonequilibrium phase transitions~\footnote{An exception is discussed in Ref.~\cite{carillo03}.}.
This mechanism (although we shall not enter into further details) is also at work in the example we shall present in the next paragraph.

\subsection{Phase transitions induced by noise: a model and a comparison between DMN and GWN}

This model corresponds to a mean-field coupling between dichotomous flows driven multiplicatively by the DMN $\xi(t)$.  The corresponding  equation
for the self-averaging stochastic variable $x(t)$ is written:
\begin{equation}
\dot{x}(t)=f(x)+g(x)\xi(t)-K(x-\langle x\rangle)\,.
\label{PT}
\end{equation}

The stationary probability density depends in an intricate way on the mean value $\langle x\rangle$, namely
\begin{eqnarray}
P_{st}(x; {\langle x \rangle})&\sim&\left(\displaystyle\frac{1}{|f_+(x;{\langle x \rangle})|}+\displaystyle\frac{1}{|f_-(x;{\langle x \rangle})|}\right)\nonumber\\
&\times&\exp\left[-k\int^x\left(\frac{dz}{f_{+}(z;{\langle x \rangle})}+
\frac{dz}{f_{-}(z;{\langle x \rangle})}\right)\right]\,,
\end{eqnarray}
where
\begin{equation}
f_{\pm}(x;{\langle x \rangle})=f(x)\pm A g(x)- K(x-{\langle x \rangle})\,.
\end{equation}
The support of $P_{st}(x;{\langle x \rangle})$
is determined by $x_{\pm}$ (with, for example, $x_- \leqslant x \leqslant x_+$), 
which are also functions of $\langle x \rangle$ determined by:
\begin{equation}
f_{\pm}(x_{\pm};{\langle x \rangle})=0\,.
\end{equation}

It follows therefore  that the self-consistency condition for the mean:
\begin{equation}
{\langle x \rangle}=\displaystyle \int_{x_-}^{x_+}\,dx\; x\; P_{st}(x;\langle x \rangle)
\end{equation}
is highly nonlinear in $\langle x \rangle$ and can thus, in principle, 
admit multiple solutions~\footnote{Just like in a mean-field like description of equilibrium situations.}.
It results the possibility of true
{\em ergodicity and symmetry-breaking phase transitions}, with the
asymptotic shape of the probability distribution dependent on the initial conditions.
In this model, the phase transition takes place between a
disordered phase with $\langle x\rangle = 0$, and an ordered phase with
$\langle x\rangle \neq 0$. These nonequilibrium phase transitions are shown to exhibit
characteristics that are similar
to second-order equilibrium phase transitions~\cite{kim98,vandenbroeck94I,vandenbroeck97I}, 
with the mean $\langle x\rangle$ playing the role of an order parameter.

For concretness,  we
shall consider the following expressions for $f(x)$ and $g(x)$:
\begin{equation}
f(x)=-x(1+x^2)^2 \quad\mbox{and} \quad g(x)=1+x^2\,.
\label{PT1}
\end{equation}
This will allow to compare this model, see Ref.~\cite{kim98}, with the situation when $\xi(t)$ is a GWN, which represents one of the first
models~\footnote{Although completely academic, this model was first introduced  because of its simplicity: no simpler forms for $f(x)$ and $g(x)$ were found to
lead to phase transitions.} introduced in the literature in order to describe a nonequilibrium noise-induced phase transition, see Refs.~\cite{vandenbroeck94I,vandenbroeck94II,vandenbroeck97I}.

\begin{figure}[h!]
\quad \vspace{1.5cm} \\
\centerline{\epsfxsize=10.1cm\epsfbox{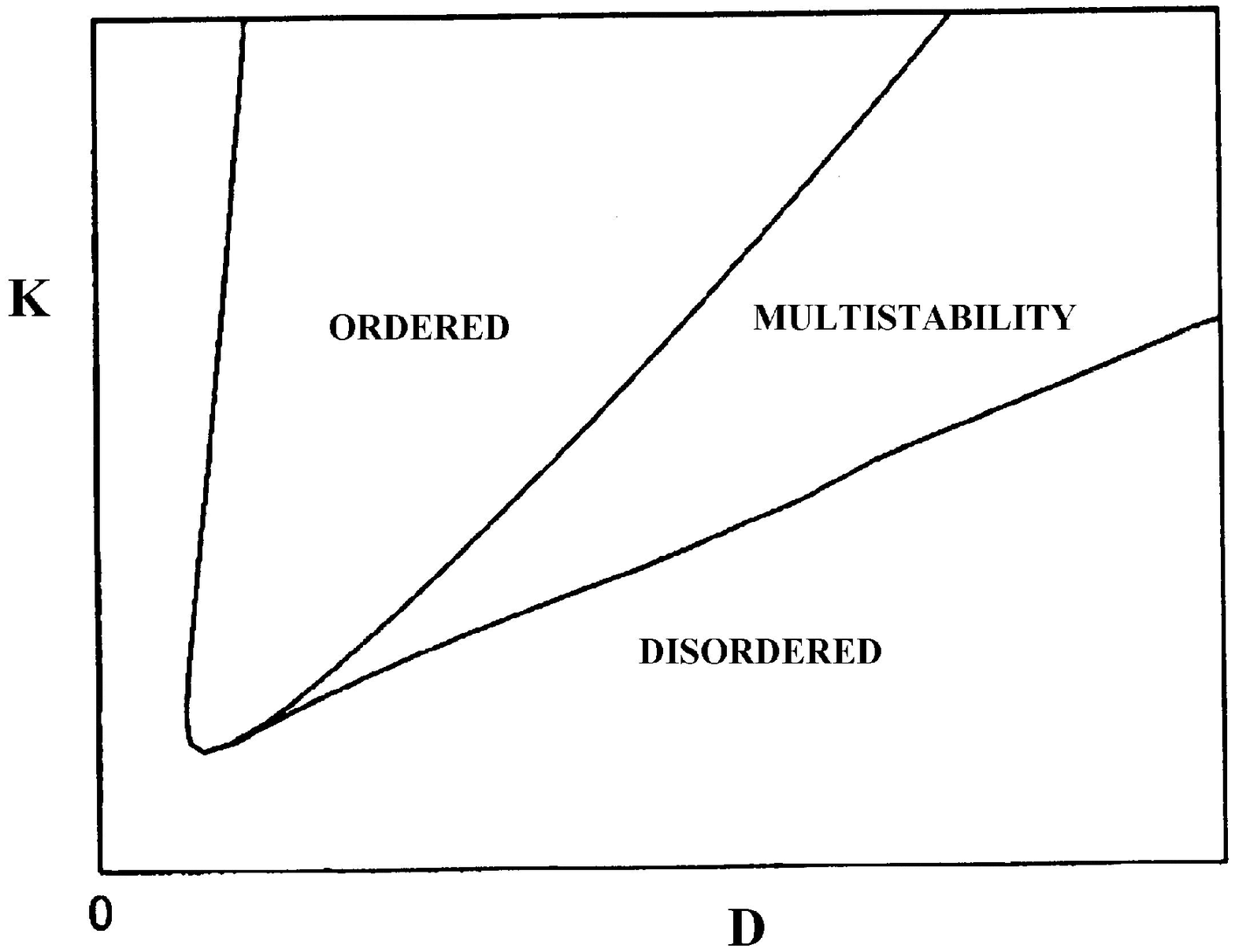}}
\hspace{-0.4cm}\centerline{\epsfxsize=10.1cm\epsfbox{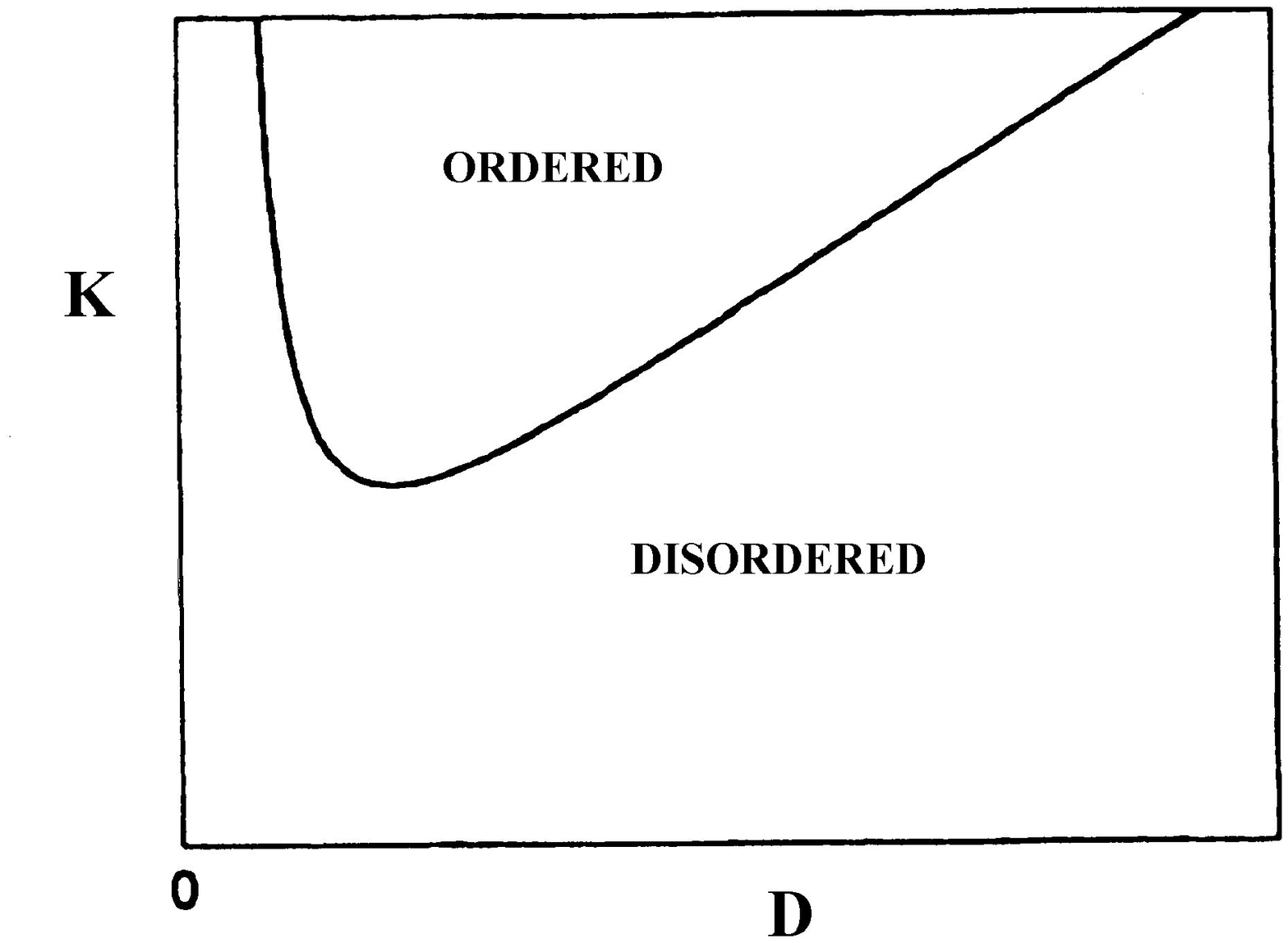}}
\caption{Qualitative phase diagram in the plane $D$ (intensity of the noise) and $K$
(coupling constant) for the mean-field models described in the main text,
equations~(\ref{PT}), (\ref{PT1}).
 Upper panel: $\xi(t)$
is a DMN with a fixed value of the transition rate $k$. Lower panel: $\xi(t)$ is a GWN.
In both cases the transition order-disorder is re-entrant with respect to the noise intensity $D$.
Note in the case of the DMN  the multistability regime, as well as the re-entrance of the
phase transition with respect to the coupling strength $K$; both effects are absent in the case of a GWN.}
\label{figure14}
\end{figure}

Following the above-cited references~\footnote{Where the phase diagram was obtained  analytically up to a point,
and afterwards  by symbolic calculations.},
in Fig.~\ref{figure14} we represented {\em qualitatively},
the phase diagram of the model in the
plane of the parameters $K$ (the coupling constant) and $D$ (the intensity of the noise),
both for the DMN ($D=A^2/2k$) and a GWN. In both cases one obtains ordered and disordered
phases and transition lines between them, and also the phenomenon of {\em re-entrance  with
respect to the intensity $D$ of the
noise.}

But, besides these, DMN also induces a {\em multistability regime} of coexistence of two stable phases,
one ordered and one disordered. This phenomenon is absent for the GWN,
i.e., it is entirely induced by the {\em finite correlation time} (colour)
of the noise. Of course, connected to this multistability, one encounters a {\em hysteretic behavior of the
order parameter $\langle x\rangle$}. Moreover, in the case of the DMN one also obtains a
{\em re-entrance with respect to the coupling strength $K$}, that does not appear
for the DMN.

So the colour of the DMN noise can alter significantly the characteristics
of the phase transition as compared to the GWN case.

\section{DICHOTOMOUS FLOWS: STATIONARY SOLUTIONS WITH UNSTABLE CRITICAL POINTS}

\subsection{Generalities}

The problem of the {\em unstable} critical points of the asymptotic dynamics was
essentially raised in the context of the existence of a {\em directed stationary flow}
(e.g., ratchet-like) in the system, $\langle \dot{x} \rangle \neq 0$.
This is a physical situation that gave rise to a lot of interest
and research, since it represents one of the paradigms of out-of-equilibrium systems:
obtaining directed motion (i.e., useful work) ``out of fluctuations",
a situation that has to be contrasted with the equilibrium one, as encoded in
the second principle of thermodynamics. The asymptotic value
$\langle \dot{x} \rangle$ is thus the quantity of main interest in this
respect.

In order to understand the peculiarities related to the existence of unstable
critical points,
let us go back to a general dichotomous flow,
\begin{equation}
\dot{x}(t)=f(x)\,+\,g(x)\,\xi(t)\,.
\end{equation}
It is clear that any systematic asymptotic drift is impossible
as long as we consider (as we did till now)
open boundary conditions and/or only stable critical points
of the alternate $f_{\pm}(x)$ dynamics.

One thus needs a {\em periodic system}:
\begin{equation}
f_{\pm}(x+L)=f_{\pm}(x)\,,
\end{equation}
with $L$ the spatial period. But this is not enough yet
in order to have a directed current. One also needs the
{\em breaking of the detailed balance}
(i.e., the fact that the system is driven
out-of-equilibrium, and thus directed current
``out-of-noise" is not ruled out by
the second principle of thermodynamics); and also
a {\em breaking of the spatial inversion symmetry}
(so that current is not ruled-out by Curie's general
symmetry principle). These conditions are discussed in detail
in the review article Ref.~\cite{reimann02},
see also Sec.~V.C below.

But when $f_{\pm}(x)$ are periodic (and at least continuous over one
spatial period), then critical points (if any) {\em appear in pairs};
therefore, the presence of stable critical points automatically
implies the presence of an equal number of {\em unstable} critical points.

The physics and the mathematics of the problem are completely different when
the asymptotic dynamics allows the ``crossing" of the unstable fixed points
(i.e., the existence of a directed current in the system) as compared to the
case when no such points are present or ``crossed".

\begin{figure}[h!]
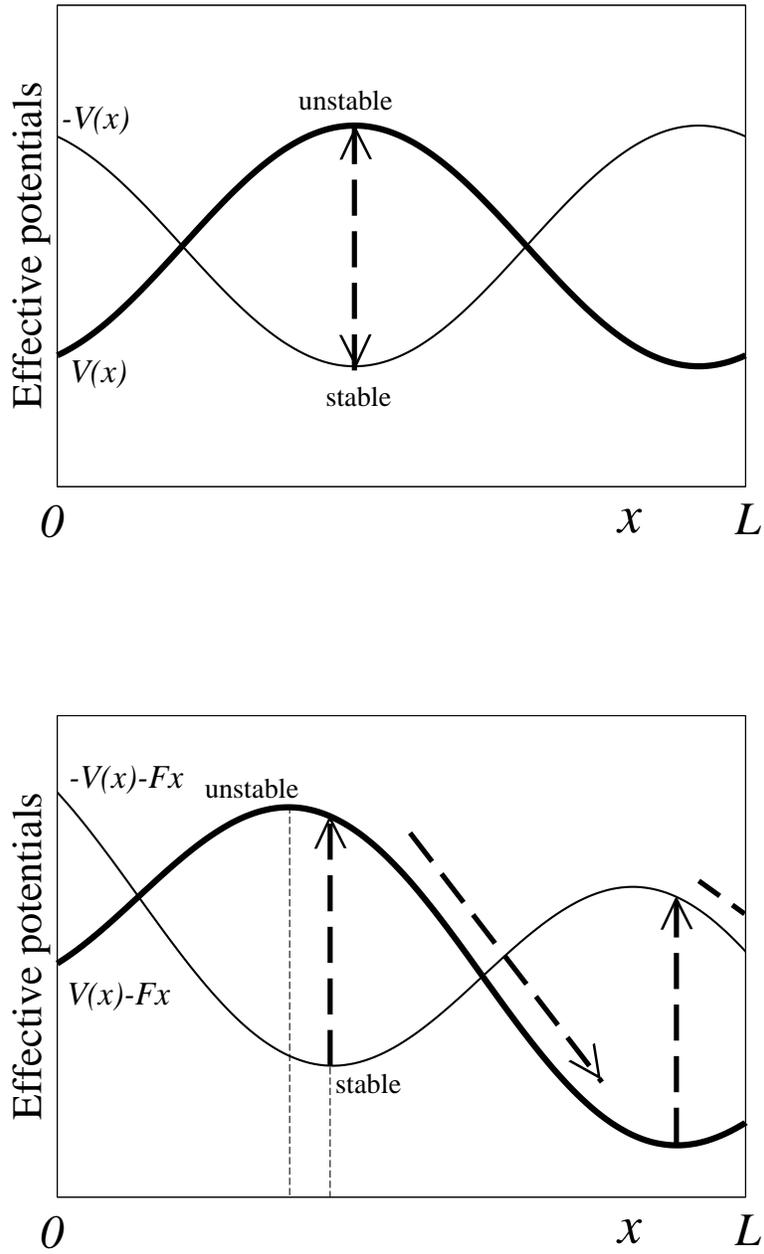

\quad \vspace{1.5cm} \\
\centerline{\epsfxsize=10cm\epsfbox{figure15a.eps}}
\quad \vspace{1.5cm} \\
\centerline{\epsfxsize=10cm\epsfbox{figure15b.eps}}
\caption{``Crossing" unstable critical points in the asymptotic dynamics
of a dichotomous flow
allows for a directed current in the system when a small symmetry-breaking force $F$
is  applied. See the main text for further details.}
\label{figure15}
\end{figure}

A very simple illustration of the changes that may appear in the
physics of the problem when unstable fixed points are ``crossed" is
given by the following example, see Fig.~\ref{figure15}. Consider an
overdamped particle in a symmetric periodic potential
$V(x)=V_0\cos(2\pi x/L+\varphi)$, that switches dichotomically
between $\pm V(x)$. In the absence of an external force, the
dynamics is restricted to the interval between two fixed points,
$[L(k\pi-\varphi)/2\pi,\,L((k+1)\pi-\varphi)/2\pi]$ ($k$ integer).
Thus, as far as the right-left symmetry is not broken, there is no
current flowing through the system, see the upper panel of the
figure. Suppose now that we apply a small external force $|F|<2\pi
V_0/L$ that breaks right-left symmetry but cannot induce ``escape"
from a finite interval in neither of the separate $f_{\pm}=\mp 2\pi
V_0/L \sin(2\pi x/L+\varphi)+F$ dynamics (corresponding to the
effective potentials $\pm V(x)-Fx$). However, switching between
these dynamics allows ``crossing" of the unstable fixed points, and
thus the appearance of running solutions with a finite average
velocity. See the lower panel of Fig.~\ref{figure15}.

From the point of view of the mathematics, the presence 
of unstable fixed points induces some serious difficulties in the
calculation of the asymptotic probability distribution function
$P_{st}(x)$.
This is the main reason why,
with a few exceptions, see Refs.~\cite{zapata98,balakrishnan01II,czernik00}
(and also Refs.~\cite{behn89,behn93} for the related problem of the mean-first passage time), the problem of dichotomous flows with unstable critical points was generally not approached. We have recently made
progress towards identifying the source of spurious divergences that arise in the
usual analytical treatment of the problem, see Refs.~\cite{bena02I,bena03,bena05},
and we are now in the position to consider this situation as well.

\subsection{Solution to the mathematical difficulties}

Consider thus a periodic dichotomous flow $f_{\pm}(x)$,
with an asymptotic state characterized by a nonzero stationary
flow $\langle \dot{x}\rangle \neq 0$ through the system.
One notices (see Ref.~\cite{bena03} for further details) that the
master equation for the stationary probability distribution function
$P_{st}(x)$:
\begin{eqnarray}
&&\left[f_+(x)f_-(x)\right]P_{st}'(x)
+\left[{(f_+f_-)'}-{(f_+f_-)}
(\mbox{ln}|f_+-f_-|)'+k(f_++f_-)\right]{P_{st}(x)}\nonumber\\
&&\nonumber\\
&&\hspace{3cm}=\frac{\langle \dot{x} \rangle}{L}
\left[2k+\frac{f_+-f_-}{2}\;\left(\frac{f_++f_-}{f_+-f_-}\right)'\;\right]
\label{master1}
\end{eqnarray}
becomes {\em singular at the critical points of $f_{\pm}(x)$}.
The crucial point of the problem resides in finding the {\em correct}
solution to this equation, i.e., the one that has {\em acceptable
mathematical and physical properties} (see below, and also Ref.~\cite{bronshtein97}).
A blind application of the
method of variation of parameters to this differential equation
leads to a solution of the form:
\begin{equation}
P_{st}(x)=\displaystyle\frac{\langle\dot{x}\rangle}{L}\;\;
\displaystyle\left|\frac{f_{+}(x)-f_{-}(x)}{f_{+}(x)f_{-}(x)}
\right|
\left[CG(x,x_0)+K(x,x_0;x)\right],
\label{case13}
\end{equation}
where $C$ is a constant of integration that arises from the general solution to
the homogeneous part of Eq.~(\ref{master1}), the second contribution is the
particular solution of the full inhomogeneous equation, $x_0$ is an arbitrary
point in $[0,L)$, and we have defined the functions:
\begin{eqnarray}
G(u,v)&=&\exp\left\{-k\int_v^udz\left[\displaystyle
\frac{1}{f_{+}(z)}+\frac{1}{f_{-}(z)}\right]\right\},
\nonumber\\
K(u,v;w)&=&\int_v^udz\;
\left[\displaystyle
\frac{2k}{f_{+}(z)-f_{-}(z)}+
\left(\frac{f_{+}(z)+f_{-}(z)}{2\left[f_{+}(z)-f_{-}(z)\right]}\right)
'\right]
\nonumber\\
&&\times \mbox{sgn}
\left[\frac{f_{+}(z)f_{-}(z)}{f_{+}(z)-f_{-}(z)}\right] \;G(w,z)
\label{case14}
\end{eqnarray}
(sgn is the signum function).

The point is that using a {\em unique} integration constant $C$ over the whole
period $[0,\,L]$ is not the right thing to do, since it leads to unphysical
divergences in the
expression of $P_{st}(x)$. More precisely, one has:\\

a). Integrable singularities at the
stable fixed points $x_s$ of $f_+$ or $f_-$:
\begin{equation}
P_{st}(x)\sim |x-x_s|^{-1+k/|f'_{(\pm)}(x_s)|} \, ,
\end{equation}
provided that the transition rate $k<|f'_{(\pm)}(x_s)|$.
Of course, these are causing no problems, and are in agreement with
the intuitive image
that there may be an accumulation of the particles at the stable
critical point of one
of the two dynamics, provided that the switch to the
alternate dynamics is not sufficiently rapid to ``sweap" particles away from
this attractor.\\

b). {\em Non-integrable singularities at the unstable fixed points}
$x_u$ of $f_+$ or $f_-$:
\begin{equation}
P_{st}\sim |x-x_u|^{-1-k/|f'_{(\pm)}(x_u)|}\,.
\end{equation}
These are, of course,
{\em meaningless and one has to find a method to avoid their appearance}.

The fundamental idea, as explained in detail in Ref.~\cite{bena03},
is then to use {\em different integration constants} in the different intervals
between fixed points, and then to {\em adjust these constants} such that
$P_{st}(x)$ acquires the right behavior, namely:
\begin{itemize}
\item $P_{st}(x)$ is continuous or  has at most integrable
singularities in $[0,\,L]$
\item The fact that the probability density is periodic:
\begin{equation}
P_{st}(x+L)=P_{st}(x)\,,
\end{equation}
with the eventual exception of some singularity points (see the simple example
below).
This condition is determined, of course, by the supposed
periodicity of the dichotomous flow.
\item The usual normalization condition one imposes
to a probability density (in this case, restricted
to a spatial period):
\begin{equation}
\displaystyle\int_0^L P_{st}(x)dx=1
\end{equation}
\end{itemize}

It can be shown that, under rather general conditions~\footnote{Actually, the
only ``stringent" condition is that $f_{\pm}(x)$ are continuous over the
spatial period. Under special circumstances, even this condition can be
relaxed, as it will becom clear from the examples we shall be considering in
the next subsections.} these requirements ensure the
{\em existence and uniquenness}
of a well-defined stationary periodic probability density.

\begin{figure}[h!]
\quad \vspace{1.5cm} \\
\centerline{\epsfxsize=11cm\epsfbox{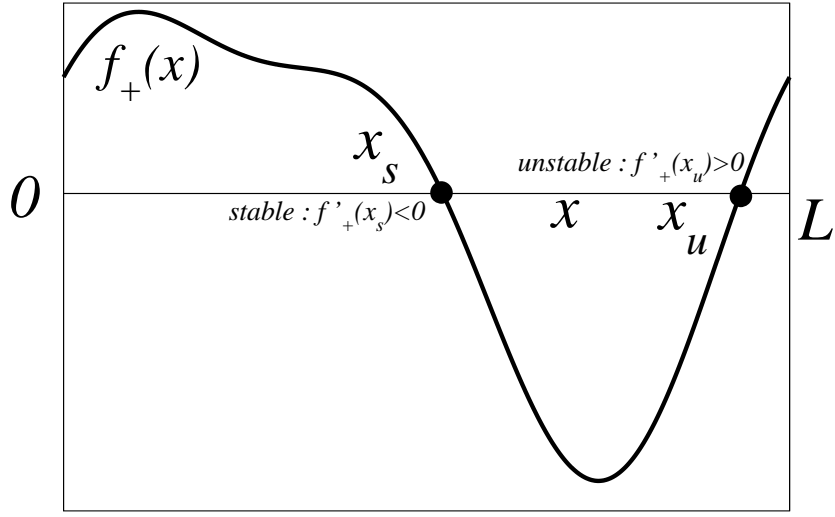}}
\caption{Qualitative representation of the ``+" velocity profile $f_+(x)=f_+(x+L)$
of a dichotomous flow,  exhibiting two critical
points: $x_s$ that is stable, and  $x_u$ that is unstable. See the main text for further details.}
\label{figure16}
\end{figure}

In order to understand the mathematical mechanism involved in
the elimination of the unphysical divergences appearing at the
unstable critical
points of the dynamics, let us consider the simplest possible situation, namely
when both $f_{\pm}(x)$ are continuous over the spatial period,
and only one of the two alternate flows has two critical points. For example,
see Fig.~\ref{figure16}, $f_{-}(x)$ has no fixed points, while $f_{+}(x)$ has a
stable $x_s$ and an unstable $x_u$ critical points (to fix ideas,
suppose $x_s<x_u$).

Consider the expression of $P_{st}(x)$ in Eq.~(\ref{case13}),
for which we shall choose {\em different integration constants}
in each of the separate
intervals $[0,x_s)$, $(x_s,x_u)$, and $(x_u,L]$ between the fixed
points.
There is {\em exactly one} choice of this constant valid for both
$(x_s,x_u)$ and
$(x_u,L)$ such that the divergence at $x_u$ is removed, namely
$C=-K(x_u,x_0; x_0)$;
and another choice valid in the
interval $[0,x_s)$ that ensures the required continuity
and periodicity of $P(x)$.
The acceptable expression for the probability density is
therefore found to be:
\begin{equation}
P_{st}(x)=
\left\{
\begin{array}{ll}
\displaystyle\frac{\langle\dot{x}\rangle}{L}\;\;
\displaystyle\left|\frac{f_{+}(x)-f_{-}(x)}{f_{+}(x)+f_{-}(x)}\right|
\left[K(L,x_u;L)G(x,0)+K(x,0;x_0)\right]
&\mbox{for}\,\,\,\,x \in [0,x_s)\\
\\
\displaystyle\frac{\langle\dot{x}\rangle}{L}\;\;
\displaystyle\left|\frac{f_{+}(x)-f_{-}(x)}{f_{+}(x)+f_{-}(x)}\right|
K(x,x_u;x),&\mbox{for}\,\,\,\,x \in (x_s,L).
\end{array}
\right.
\label{case110}
\end{equation}

These expressions can be further simplified if one takes as the basic
period not
$[0,L]$, but $[x_s,x_s+L]$. Then the simple-looking, ``compact"
expression
\begin{equation}
P_{st}(x)=\displaystyle\frac{\langle\dot{x}\rangle}{L}\;\;\displaystyle
\left|\frac{f_{+}(x)-f_{-}(x)}{f_{+}(x)f_{-}(x)}\right|
K(x,x_u;x)
\label{compact}
\end{equation}
holds throughout
this new basic period. Moreover, $P_{st}(x)$ meets all the requirements
enumerated above. In particular, let us check its behavior at the fixed points
$x_{1,2}$. For $x=x_u$ (the unstable fixed point that caused problems before!),
the above expression of $P_{st}(x)$ presents an indeterminacy of the type
``0/0": by applying {\em H\^ospital's rule} one simply finds that  $P_{st}(x)$
is {\em continuous at the unstable fixed point}:
\begin{equation}
\lim_{x\rightarrow x_u}P_{st}(x)=
\displaystyle\frac{\langle\dot{x}\rangle}{L}\;
\displaystyle\frac{
2k/f'_{+}(x_u)+1}
{f_{-}(x_u)\left[k/f'_{+}(x_u)+1\right]}\,.
\end{equation}
The behavior at the {\em stable} fixed point $x_s$  is continuous
\begin{equation}
\lim_{x\rightarrow x_s} P_{st}(x)=
\displaystyle\frac{\langle\dot{x}\rangle}{L}\;
\displaystyle
\frac{2k/|f'_{+}(x_s)|-1}
{f_{-}(x_s)\left[k/|f'_{+}(x_s)|-1\right]}
\end{equation}
for $k/|f'_{+}(x_s)|>1$, while it has a power-law integrable divergence for
$k/|f'_{+}(x_s)|<1$, and a marginal logarithmic divergence for
$k/|f'_{+}(x_s)|=1$. One also notices the periodicity of $P_{st}(x)$
(except at the eventual singularity point $x_s$).

Imposing the normalization of $P_{st}(x)$ over one spatial period, one obtains
finally the expression of the {\em asymptotic mean velocity} through the
system, which is the most important quantity for most practical purposes:
\begin{eqnarray}
\langle\dot{x}\rangle&=&L\left\{
\int_{x_s}^{x_s+L}dx\displaystyle
\left|\displaystyle\frac{f_{+}(x)-f_{-}(x)}{f_{+}(x)f_{-}(x)}\right|
\;\;\displaystyle\int_{x_u}^{x}dz\;\mbox{sgn}\left[
\displaystyle\frac{f_{+}(z)f_{-}(z)}{f_{+}(z)-f_{-}(z)}\right]
\right.\nonumber\\
&&\hspace{-1cm}\times \left.
\left[\displaystyle\frac{2k}{f_{+}(z)-f_{-}(z)}+
\displaystyle\left(\frac{f_{+}(z)+f_{-}(z)}{2\left[f_{+}(z)-f_{-}(z)
\right]}\right)^{'}\right]\exp\left[-k\int_z^x dw\;
\left(\frac{1}{f_{+}(w)}+\frac{1}{f_{-}(w)}\right)\right]
\right\}^{-1}.
\nonumber\\
&&
\label{asymptotic}
\end{eqnarray}

This mathematical reasoning can be generalized to any type of combination of
critical points of the alternate dynamics of the dichotomous flow, as shown in
detail in the above-mentioned references, and we shall illustrate it on three
examples below.

The first one refers to an interesting physical situation,
namely that of {\em hypersensitive response}, that is actually in
tight connection with the system already described qualitatively
in Fig.~\ref{figure15}. The second one referes to a {\em rocking ratchet},
and the third one is an important particular realization of such a
ratchet, namely a {\em stochastic Stokes' drift} effect.

\subsection{Nonlinear and hypersensitive response with DMN}

Generally and somehow loosely speaking,
the term {\em hypersensitive response} referes to a
large and highly-nonlinear response of a noisy,
out-of-equilibrium system to a
small, directed external forcing. This phenomenon
was discovered rather recently, and a lot of
attention was given to several of its variants, both
theoretically, see
Refs.~\cite{tarlie98,berdi98,ginzburg98,ginzburg99I,ginzburg99II,ginzburg01,geraschenko01,mankin03}, and experimentally,
see Refs.~\cite{geraschenko98,geraschenko00,ginzburg02,fateev02,ginzburg03}.

We have recently introduced a simple model of dichotomous flow that exhibits
such a hypersensitive behavior, see Refs.~\cite{bena02I} and \cite{bena05}. 
It describes an overdamped particle that switches dichotomically between a
symmetric potential $V(x)$ and its negative, as represented schematically in
Fig.~\ref{figure15}. When the symmetry of the system is slightly broken by a
small directed external force $F$, the system responds with a nonlinear
drift~\footnote{Although very simple and seemingly robust, this particular
system has not been
realized experimentally yet. We are currently investigating the possibility of
its realization using a SQUID.}.
The dynamics can be described by the following stochastic equation (with the
DMN acting multiplicatively):
\begin{equation}
\dot{x}=F+\xi(t)v(x)\,,
\label{hyper1}
\end{equation}
where $v(x)=-V'(x)$, with $v(x)=v(x+L)$,
and (through a rescaling of $v(x)$) the DMN $\xi(t)=\pm 1$.\\

(a) When the external force is sufficiently large, {\em there is no fixed point
in either $f_{\pm}(x)$ dynamics} [$F^2-v^2(x)\neq 0$ for any $x\in [0,L)$]. By
integrating the stationary master equation~(\ref{master1})
corresponding to Eq.~(\ref{hyper1}),
one obtains the following expression for the stationary probability density,
\begin{equation}
P_{st}(x)=
\frac{\langle\dot{x}\rangle}{LF}
\left\{1 + \frac{v(x)\displaystyle\int_x^{x+L}dz\;
v'(z)\exp\left[-\int_z^x dw \frac{2kF}
{F^2-v^2(w)}\right]}{[F^2-v^2(x)]
\left\{\exp\left[\displaystyle\int_0^L
dz \frac{2kF}{F^2-v^2(z)}\right]-1
\right\}}
\right\}\,.
\label{hyper3}
\end{equation}

By imposing the normalization condition of $P_{st}(x)$ over $[0,\,L)$,
one obtains finally the asymptotic mean velocity:
\begin{equation}
\frac{\langle\dot{x}\rangle}{F}=
\left\{1+\frac{\displaystyle\int_0^Ldx\,\frac{v(x)}{F^2-v^2(x)}
\int_x^{x+L}dz\;v'(z)\exp\left[-\int_z^x
dw \frac{2kF}{F^2-v^2(w)}\right]}
{L\left\{\exp\left[\displaystyle\int_0^L dz \frac{2kF}{F^2-v^2(z)}
\right]-1\right\}}\right\}^{-1}\,.
\label{hyper4}
\end{equation}\quad \\

(b) Let us consider the more interesting case of {\em small forcing},
when the {\em asymptotic dynamics has critical points}. In order to fix ideas
(and without losing any relevant element), we take $v(x)$ continuously
decreasing in $[0,\,L/2]$ and symmetric about $L/2$, $v(x+L/2)=-v(x)$.
Then $P_{st}(x+L/2)=P_{st}(x)$, which allows us to concentrate on a
half-period.  In this case, the ``--" dynamics has an unstable fixed point
$x_u$, and the ``+" dynamics has a stable critical point $x_s$ (with $x_s >x_u$)
in $[0,\,L/2]$. According to the general discussion in the preceding section,
one obtains the following physically acceptable solution in
the interval $[x_s-L/2, \,x_s]$ (that can be afterwards extended by periodicity
to the whole real axis):
\begin{eqnarray}
P_{st}(x) &=&
\displaystyle\frac{\langle\dot{x}\rangle}{LF}
\left\{1+\displaystyle\frac{v(x)}{|F^2-v^2(x)|}
\displaystyle\int_{x_u}^xdz\;
\mbox{sgn}\left[F^2-v^2(z)\right]v'(z)
\right. \nonumber\\
&&\nonumber\\
&&\hspace{3cm}\times\,
\left.\exp\left[-\displaystyle\int_z^x
dw\frac{2kF}{F^2-v^2(w)}\right]\right\}\,.\label{hyper6}
\end{eqnarray}
At the unstable fixed point $x_u$, the probability
density is continuous,
\begin{equation}
\lim_{x\searrow x_u}P_{st}(x)
=\lim_{x\nearrow x_u}P_{st}(x)
=\frac{\langle\dot{x}\rangle}
{LF\left\{1-\displaystyle\frac{1}{2(k/|v'(x_u)|+1)}\right\}}\,.
\end{equation}
At the stable fixed point $x_s$, depending on the transition rate $k$, the
probability density is either continuous,
\begin{equation}
\lim_{x\nearrow x_s}P_{st}(x)
=\lim_{x \searrow (x_s-L/2)}P_{st}(x)
=\frac{\langle\dot{x}\rangle}{LF\left\{1+\displaystyle\frac{1}{2(k/|v'(x_s)|-1)}\right\}}
\end{equation}
(for $k/|v'(x_s)|>1$), or divergent but integrable
(for $k/|v'(x_s)| \leqslant 1$).
It is the presence of these fixed points, and in particular the divergences of
$P_{st}(x)$ that cause a highly nonlinear conductivity of the system, as discussed below.

Finally, from the normalization of $P_{st}(x)$, the average
asymptotic velocity is obtained as
\begin{eqnarray}
\frac{\langle\dot{x}\rangle}{F}&=&
\left\{1+\frac{2}{L}\displaystyle\int_{x_s-L/2}^{x_s}dx\;
\frac{v(x)}{|F^2-v^2(x)|}\right.
\displaystyle\int_{x_u}^xdz\;
\mbox{sgn}\left[F^2-v^2(z)\right]v'(z)\nonumber\\
&&\nonumber\\
&&\hspace{5cm}\times\,\left.
\exp\left[-\displaystyle\int_z^xdw\frac{2kF}{F^2-v^2(w)}
\right]\right\}^{-1}\,.
\label{hyper11}
\end{eqnarray}

\begin{figure}[h!]
\quad \vspace{1.5cm} \\
\centerline{\epsfxsize=12.cm\epsfbox{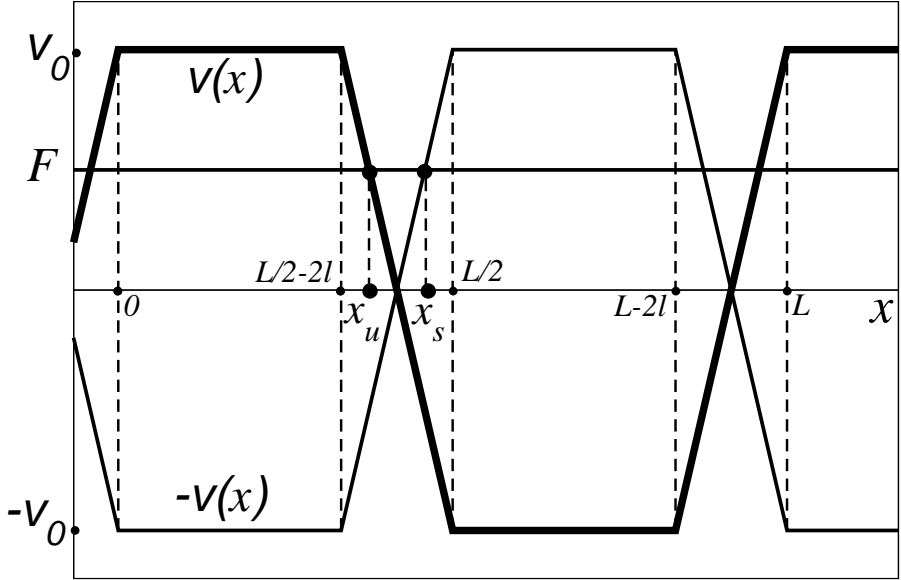}}
\caption{Piecewise linear profile $v(x)$ for the hypersensitive response model,
see Eq.~(\ref{hyper1}) in the
main text. A small applied force $F$  induces critical points in both ``+" and ``--" dynamics
($x_u$ is an unstable fixed point of the ``--" dynamics, while $x_s$ is a stable fixed point of the ``+" dynamics). Note the symmetry of the system with respect to $x=L/2$.}
\label{figure17}
\end{figure}

\begin{figure}[h!]
\quad \vspace{1.5cm} \\
\centerline{\epsfxsize=12.cm\epsfbox{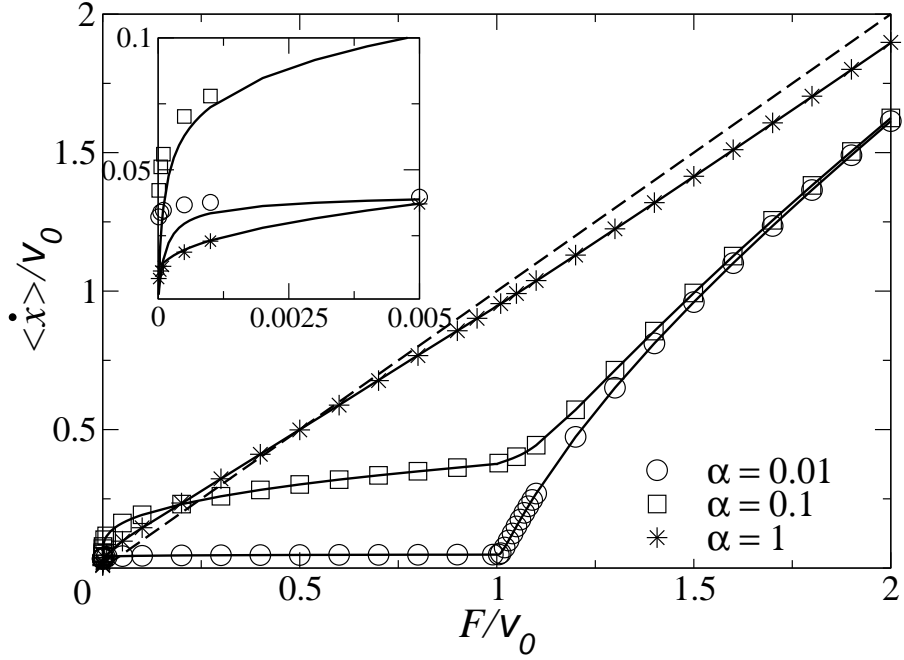}}
\caption{The mean asymptotic velocity $\langle \dot{x}\rangle/v_0$
as a function of the external driving $F/v_0$ for different values of the
control parameter $\alpha=lk/v_0$ (for the meaning of $l$ and $v_0$, see Fig.~\ref{figure17}). The solid lines are the corresponding
result of the theory, and the
symbols represent the result of  averaging over
numerical simulations of the dynamics of an ensemble of 20000 particles.
The dashed line corresponds to the linear response $\langle \dot{x}\rangle=F$.
There are four different regimes
for the response of the system to the external forcing $F/v_0$, see the main text.
The inset is a detail of the region around the origin (corresponding to the hyper-nonlinear regime).
We are grateful to Prof. R. Kawai for providing us the data for this plot.
See also Fig.~5 in Ref.~\cite{bena05}.}
\label{figure18}
\end{figure}

These results for $P_{st}$ and the mean current are general and exact.
However, they still involve triple integrals, and are too complicated to offer a
picture of what is going on in the system. We shall thus consider further the
particular case of a {\em piecewise profile of $v(x)$}, as represented in
Fig.~\ref{figure17}.
In this case, the integrals can be evaluated explicitely, and it is found that
the behavior of the system is extremely rich.
The response $\langle \dot{x}\rangle$ to the external perturbation
$F$ is highly-dependent, in a non-monotonic way,
on the transition rate $k$ of the DMN,
more precisely on the control parameter $\alpha=lk/v_0$
(for the significance of $l$ and $v_0$ refer to Fig.~\ref{figure17}).
One notices the existence of four different regimes for the response,
as illustrated in Fig.~\ref{figure18}. \\

(a) A first trivial {\em linear response} regime corresponds
to very high applied forces
$F/v_0 \gg 1$, when the details of the substrate potential $\pm  V(x)$
are ``forgotten", and $\langle \dot{x}\rangle\approx F$. All curves in
Fig.~\ref{figure18} approach this limit with increasing $F$.\\

(b) A second {\em linear response} regime $\langle \dot{x}\rangle\approx F$
appears when $\alpha \gg 1$, i.e.,
for very high transition rate of the DMN, such that the effects of the fluctuating
forces are smeared out. This regime is visible on Fig.~\ref{figure18} for $\alpha=1$.\\

In none of these regime is the response dependent on the characteristics of the DMN.
The following two regimes are, on the contrary, far from trivial and appear only
in the presence of critical points of the alternate $f_{\pm}(x)$ dynamics.\\

(c) The {\em adiabatic regime}  of constant response
appears for very slow switching rates $k$, more precisely when in-between two flips of
the potenatial (refer to Fig.~\ref{figure15}) the particles have enough time to move
between two succesive extrema of the potential, and eventually to wait
in a minimum of the potential till the next flip, that will put them in motion again.
Therefore, the condition for this regime
is that the average time between switches $k^{-1}$ is much longer than the typical escape
time from a region close to a maximum of the potential, $\tau\sim -(l/v_0) \mbox{ln}f$;
therefore: $1>f \gg \exp(-1/\alpha)$.
Then the mean velocity is simply half of the spatial period of the substrate potential
divided by the mean switching time,
(i.e., it is independent of the
applied force $F$ and is directly proportional to $k$):
\begin{equation}
\langle \dot{x}\rangle\approx lk/v_0\,.
\end{equation}
This regime is well seen on Fig.~\ref{figure18} for
$\alpha=0.01$.\\

(d) Finally, the {\em hyper-nonlinear} regime which is realized for small forcing,
$f <\exp(-1/\alpha)\ll 1$. In this case, the particles manage to advance to the next
minimum of the potential only in the exponentially rare cases when the DMN persists
in the same state for sufficiently long time, much longer than both
$k^{-1}$ and $\tau$ (see again Fig.~\ref{figure15} and see above for the meaning of $\tau$).
The mean velocity then falls
rapidly to zero with decreasing $F$:
\begin{equation}
\langle \dot{x}\rangle \approx \frac{v_0^2 L}{8 l^2 k\;\mbox{ln}^2(F/v_0)}\,,
\end{equation}
i.e., it is inversely proportional to $k$, and the corresponding
diverging susceptibility
\begin{equation}
\chi=\displaystyle \left.\frac{\partial \langle\dot{x}\rangle}{\partial F}\right|_{F=0} \rightarrow \infty
\end{equation}
indicates the highly-nonlinear and sensitive character of the response in this region.
This regime appears for all the values of $\alpha$ and sufficiently small $F$,
as seen on the inset of Fig.~\ref{figure18}.

We are therefore again in a situation that is contrary to the intuition and
to what is usually encountered in
equilibrium systems, namely {\em a strongly nonlinear
response for small forcing,
and a linear response for large forcing}.

\subsection{Ratchets with DMN}

One of the paradigms of out-of-equilibrium system
(with an overwhelming literature these last years) is the {\em ratchet effect}.
As it was already briefly mentioned above, a {\em ratchet} is,
roughly speaking, a device
that allows to get work (i.e., directed transport) out of fluctuations.
Although one can think of macroscopic ratchets
(e.g., self-winding wrist-watches,
wind-mills), more interesting from the conceptual point of view are
the {\em microscopic rectifiers}, for which microscopic thermal
fluctuations are relevant. Indeed, while
the second law of
thermodynamics rules out directed transport (apart from transients) in a
spatially-periodic system in contact with a single heat bath~\footnote{A
classical illustration is given by the Smoluchowski-Feynman ratchet-and-pawl
device in contact with a single thermostat, see Ref.~\cite{feynman}.}, 
there is no such fundamental law that prohibits stationary
directed transport in a system driven  out-of-equilibrium by a
deterministic or stochastic forcing.
Such a driving forcing can be provided, for example,
by a DMN that can act either multiplicatively, or additively.

Besides the {\em breaking of detailed
balance}, a further indispensable requirement for directed transport
is {\em breaking of spatial symmetry}~\footnote{Such that directed current is
not ruled out by Curie's symmetry principle, ``If a certain phenomenon is not
ruled-out by symmetries, then it may/will appear".},
see Ref.~\cite{reimann02}.
There are three main possible mechanisms of symmetry-breaking, namely
(i) a built-in asymmetry of the system (in the absence of the driving
perturbation); (ii) an asymmetry induced by the perturbation; (iii) a
dynamically-induced asymmetry, as a collective effect,
through an out-of-equilibrium symmetry-breaking
phase transition.

The case of an out-of-equilibrium driving by a {\em multiplicative} DMN
corresponds to the so-called {\em flashing ratchet}:
generically, an overdamped Brownian particle) jumps dichotomously,
at random, between two asymmetric periodic potentials $V_{\pm}(x)$:
\begin{equation}
\dot{x}(t)=-V'(x)[1+\xi(t)]+\xi_{GWN}(t)\,,
\end{equation}
where $\xi(t)$ is the DMN $\pm A$ with transition rate $k$, $\xi_{GWN}$ is a Gaussian white noise, and, of course, $V_{\pm}(x)=V(x)(1\pm A)$.
The particular case of $V_{-}(x)=0$, i.e., a flat potential corresponding
to a free diffusion in the ``--" dynamics, is called {\em on-off ratchet}.
This results in a net flow of the particles,
in a direction that is determined by the asymmetry of $V(x)$.
Various models, aspects, and experimental realizations of flashing or on-off ratchets,
including sometimes the effect of inertia,
were discussed in Refs.~\cite{bug87,ajdari92,astumian94,prost94,rousselet94,chauwin95,faucheux95,mielke95,gore97,kula98I,chen99}, see also Ref.~\cite{reimann02}.

Systems driven out of equilibrium by an {\em additive} DMN belong to the
class of the so-called {\em rocking ratchets}:
an asymmetric basic potential $V(x)$ is rocked  by a zero-mean additive force
(a DMN $\xi(t)$ in the cases of interest to us):
 \begin{equation}
\dot{x}(t)=-V'(x)+\xi(t) +\mbox {eventually}\; \xi_{GWN}(t)\,.
 \end{equation}
This leads generically to an
asymmetry in the nonlinear response of the system, and thus to a
systematic (directed) motion, be it in the presence or in the absence of a thermal noise.
Inertia of particles was also find to have an important effect on the direction of the drift.
The literature on various model-realizations and applications is huge, and we cite here
only a few references,~\cite{magnasco93,svoboda93,doering94,astumian94,chialvo95,mielke95,millonas96,kula96,zapata96,berdichevsky97,park97,zapata98,kula98,kula98I,nikitin98,arizmendi98,li98},
and also Ref.~\cite{reimann02} for further examples.

\begin{figure}[h!]
\quad \vspace{1.5cm} \\
\centerline{\epsfxsize=11cm\epsfbox{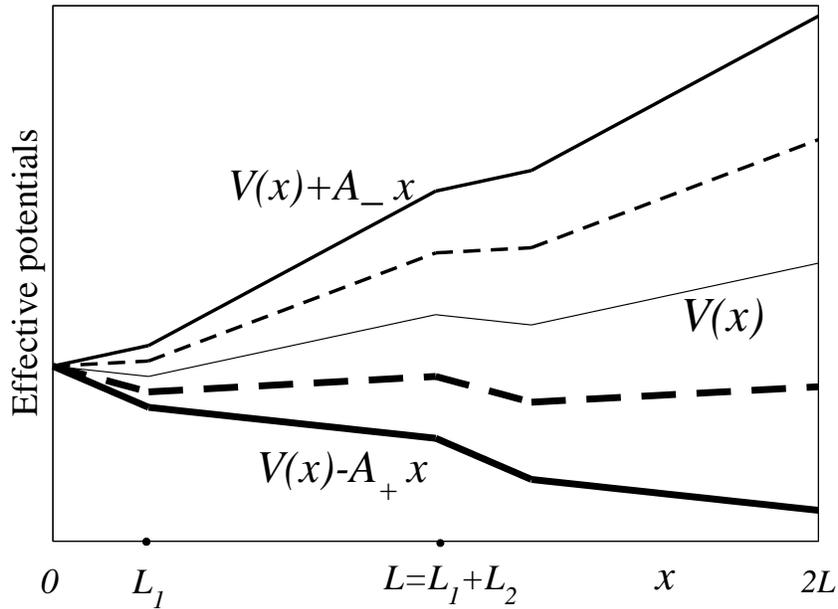}}
\caption{Effective potentials for the rocking-ratchet model in Eq.~(\ref{rock1}). Depending on the values $\pm A_{\pm}$ of the DMN forcing, one may have running solutions in both $f_{\pm}$ dynamics (the solid lines $V(x)\mp A_{\pm}x$), or running solutions in only one of the alternate dynamics (dashed lines), or no running solutions at all. We also represented
the unperturbed basic potential $V(x)$.}
\label{figure19}
\end{figure}

As an illustration, we propose here a very simple analytically solvable model
of a {\em rocking ratchet}, see Ref.~\cite{bena03} for a detailed discussion.
It is described by the following stochastic differential equation:
\begin{equation}
\dot{x}(t)=f(x)+\xi(t)\,,
\label{rock1}
\end{equation}
with an {\em additive and asymmetric} DMN $\xi(t)$,
that takes the values $\pm A_{\pm}$ with transition rate $k$ between these states.
In order to fix ideas, we shall suppose $A_- > A_+ >0$.
Here
\begin{equation}
f(x)=-V'(x)= \left\{
\begin{array}{ll}
+f_1\,,&x\in[0,\,L_1)\\
-f_2\,, &x\in[L_1,L_1+L_2)
\end{array}
\right.
\end{equation}
with $f(x)=f(x+L)\,,\quad L=L_1+L_2$, and  $0<f_2<f_1$.
This corresponds to an overdamped particle gliding in a rocked sawtooth potential,
as depicted qualitatively in Fig.~\ref{figure19}.

It is obvious that three distinct situations are possible, depending on the values
$\pm A_{\pm}$ of the DMN. In all these cases one can obtain closed analytical expressions for the stationary probability density $P_{st}(x)$ and the mean asymptotic velocity $\langle \dot{x}\rangle$. We refer the reader to Ref.~\cite{bena03} for the exact results, while here we shall comment on a few features. One can have: \\

(a) A regime of {\em strong forcing}, when there is no critical point in any of the
alternate $f_{\pm}(x)$ dynamics, and running solutions appear for both tilts.
With our conventions, this corresponds to $A_->f_1$ and $A_+>f_2$.
There results a non-zero current through the system, determined by the interplay between the characteristics of the noise and those of the basic potential $V(x)$ (in particular, its bias, if any).

One can also consider the  GWN limit, for which already known results~\cite{risken84} are recovered
in a very simple way. In particular, the current through the system is strictly detemined by the bias of $V(x)$, such that, when the deterministic potential is unbiased, one recovers
the equilibrium state with a Boltzmann distribution and no current.
\\

(b) The regime of {\em intermediate forcing}, when there are critical points in only one
of the alternate dynamics. In our case, this happens when $A_->f_1$ and $A_+<f_2$,
and only the ``+" dynamics has critical points.
It is obvious that the sign of the flow is determined by that of the dynamics without fixed points (``--" in our case);  sometimes  the current may thus  be  opposite to
the bias of the basic potential $V(x)$.

One notices the possibility of
current reversal when varying the amplitude of the of the noise
(at fixed $k$), i.e., when passing from regime (a) to regime (b).

As a limiting case of both (a) and (b) regimes, one can consider the white
shot-noise limit (see Sec.~I.A), and recovers easily the results of Ref.~\cite{czernik00}.\\

(c) Finally, the regime of {\em weak forcing}, when both alternate dynamics have critical points. There is not too much interest in it, since there is no flow through the system.
Additional thermal noise is needed to generate rectified motion, and this problem has been addressed (mainly in the case of adiabatically slow forcing)  in Refs.~\cite{magnasco93,astumian94,doering94,kula96,kula98,luczka97}.

\subsection{Stochastic Stokes' drift}

\begin{figure}[h!]
\quad \vspace{1.5cm} \\
\centerline{\epsfxsize=10.cm\epsfbox{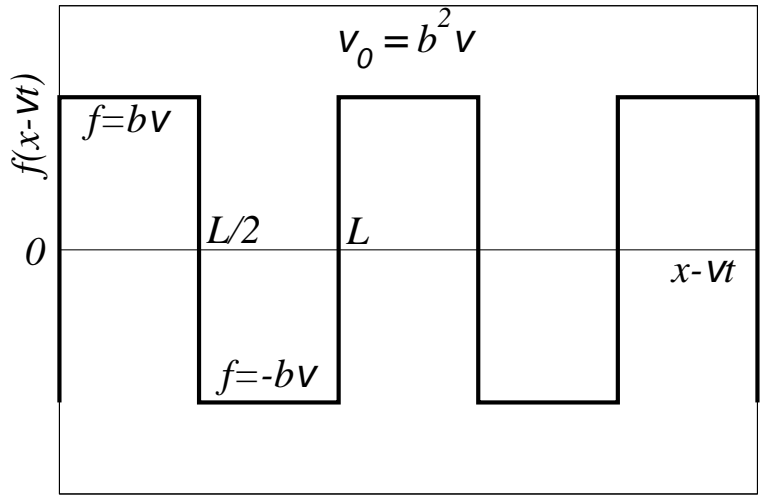}}
\caption{Schematic representation of the deterministic Stokes' drift for an overdamped particle driven by a longitudinal square-like wave, see the main text.}
\label{figure20}
\end{figure}

{\em Stokes' drift} refers to the systematic motion that a tracer acquires in a viscous fluid under the action of a longitudinal wave traveling through this fluid,
see the original reference~\cite{stokes47}. The {\em deterministic effect} (that does not account for the fluctuations or perturbations in the system) has a simple intuitive explanation, as illustrated through the example of Fig.~\ref{figure20}~\cite{vandenbroeck99}.
Consider a longitudinal square-like wave of wavelength $L$,
propagating with velocity $v$, and an overdamped
tracer particle, that is entrained with a force $f=\dot{x}=bv$ while in the
crest part of the wave, and $f=\dot{x}=-bv$ while in the trough part (with $0<b<1$ the situation of physical relevance). The suspended particle spends a longer time in the regions of the wave train where it is driven in the direction of propagation of the wave, namely 
 $t=L/[2(1-b)v]$,  than the time it spends in those regions where the drag force acts in the opposite direction, namely   $t=L/[2(1+b)v]$. Therefore, the particle is driven on the average in the direction of the wave propagation, with $v_0=b^2v$, the deterministic value of the Stokes' drift.
This effect has been studied in various practical contexts, ranging from the motion of tracers in meteorology and oceanography, to the doping impurities in crystal growth, see
the citations in Refs.~\cite{bena00,bena05}.

Recent studies, see Refs.~\cite{janson98,vandenbroeck99,bena00,bena05}
and references therein, show the importance the {\em stochastic effects}
may have on Stokes' drift.  The thermal diffusion of the dragged particles,
as well as the application of a coloured external perturbation modify markedly both the {\em direction} and the {\em magnitude} of the drift velocity.

We introduced~\cite{bena00,bena05} a very simple,
analytically tractable  model for a stochastic Stokes' drift,
described by a dichotomous flow as:
\begin{equation}
\dot{x}(t)=f(x-vt) +\xi(t)\,.
\end{equation}
Here $f(x-vt)$ corresponds to the block-wave represented in Fig.~\ref{figure20},
and $\xi(t)$ is a symmetric DMN of values $\pm A$ and transition rate $k$.
One can perform a simple transformation of variables
(which corresponds to going to the wave co-moving frame),
$y=x-vt$, through which the model can be mapped onto a {\em rocking ratchet} problem (as described in the previous section, with an asymmetric basic sawtooth potential and an additive DMN):
\begin{equation}
\dot{y}(t)=F(y)+\xi(t)\,,
\end{equation}
with $F(y)=-(1-b)v$ for $y \in[0,\,L/2)$ and $F(y)=-(1+b)v$ for $y \in [L/2,\,L)$.

As discussed previously, the behavior
of the system, in particular the solution of the associated master equation for the stationary probability density $P_{st}(x)$ and the asymptotic drift velocity $\langle \dot{y}  \rangle =
\langle \dot{x}\rangle -v$, depend on whether or not there are critical
points in the dichotomous dynamics.
There are two important effects that appear due to the stochastic DMN forcing:
(i) {\em the enhancement of the drift as compared to its deterministic value}; (ii)
the possibility of {\em drift reversal} when modifying the amplitude of he noise.
Refer  to~\cite{bena00,bena05} for the detailed calculations.

\begin{figure}[h!]
\quad \vspace{1.5cm} \\
\centerline{\epsfxsize=11.cm\epsfbox{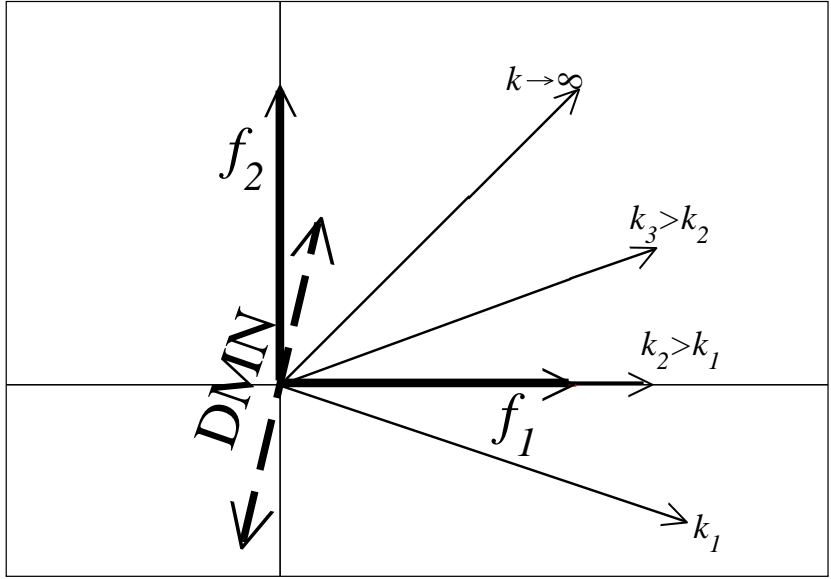}}
\caption{Schematic representation of the nonadditive character of the stochastic Stokes' drift.
Consider an overdamped tracer under the action of two identical square waves $f_1$ and $f_2$ (thick lines), propagating in orthogonal directions, as well as a DMN
(thick dashed line). DMN is taking the values $\pm A$ and is acting along a direction that
does not coincide with the bisecting line of the two waves $f_1$ and $f_1$.
The direction and the magnitude of the resulting drift velocity
(thin lines) change
significantly with the transition rate $k$ of the DMN. $k=\mbox{infinity}$ corresponds to the deterministic, noiseless limit of linear superposition of Stokes' drift.}
\label{figure21}
\end{figure}

Also, noise induces a nonlinear dependence of the drift
on the amplitude of the wave. Therefore, if several waves
are present, their contributions {\em are not additive},
which is a generic feature of stochastic Stokes' drift,
contrary to its deterministic counterpart.
In particular, as illustrated qualitatively in Fig.~\ref{figure21},
if two orthogonally-propagating but otherwise identical waves are present,
one can induce a significant change in the direction and magnitude
of the resulting drift simply by changing the transition rate
$k$ of the DMN. This is an effect that may have
important practical applications, e.g., in directing
doping impurities in crystal growth.

\section{ESCAPE PROBLEMS FOR SYSTEMS DRIVEN BY DMN}

\subsection{Mean first-passage times for dichotomous flows}

The mean first-passage time (MFPT) represents the mean value of the
time at which a stochastic variable, starting at a given initial value, reaches
a pre-assigned threshold value for the first time. It is a concept with many
applications in physics, chemistry, engineering -- ranging from the decaying
of metastable and unstable states, nucleation processes,
chemical reaction rates,
neuron dynamics, self-organized criticality, dynamics of spin systems,
diffusion-limited aggregation, general stochastic systems with absorbing states,
etc., as discussed
in Ref.~\cite{redner01}.

MFPT expressions were obtained in closed analytical form for some
particular Markovian and non-Markovian
stochastic processes, including one-dimensional Fokker-Planck
processes, continuous-time random walks or persistent random walks
with nearest and next-to-nearest neighbour jumps, and birth-and-death processes, see Refs.~\cite{stratonovich63,hanggi81,weiss81,bala83,chris86,hanggi90}
for some examples.

For general non-Markovian processes the problem of the MFPT is delicate and intricate, as spelled out first in Ref.~\cite{hanggi83}. However,
for general dichotomous flows one can obtain
exact results, using various approaches and techniques
(backward equations, stochastic trajectory counting and analysis), see
Refs.~\cite{hanggi85,sancho85,rodriguez86,masoliver86I,masoliver86II,masoliver86III,doering87,weiss87,tsironis88,balakrishnan88II,behn89,kus91,behn93,olarrea95I,olarrea95II}.

Consider an overdamped particle that starts at $t=0$ in $x_0$
inside some interval  $[x_A, \,x_B]$ of the real axis,
and being driven by the dichotomous flow
$\dot{x}(t)=f(x)+g(x)\xi(t)$.
Denote by $T_{\pm}(x_0)$  the MFPT corresponding to the particle leaving the
interval   $[x_A, \,x_B]$, provided that the initial value of the DMN $\xi(t=0)$
was $\pm A$, respectively.
The coupled equations for $T_{\pm}(x_0)$ are readily found to be:
\begin{eqnarray}
&&\left[f(x_0)+Ag(x_0)\right]\,\frac{dT_+(x_0)}{dx_0}-k(T_+-T_-)=-1\,,\\
&&\left[f(x_0)-Ag(x_0)\right]\,\frac{dT_-(x_0)}{dx_0}-k(T_--T_+)=-1\,.
\end{eqnarray}
The true difficulty here consists in  assigning the correct
{\em boundary conditions},
corresponding either to absorbing or to  (instantaneously) reflecting boundaries,
and also in treating the critical points of the alternate ``+" and ``--" dynamics. For further details see the above-cited references.

The particular case of the so-called {\em bistability driven by} DMN,
when $f(x)$ derives from a bistable potential $V(x)$, $f(x)=-V'(x)$,
received a lot of attention, see Refs.~\cite{hanggi83I,chris84,balakrishnan88II,heureux88,heureux89,porra91,porra92,heureux95}. In these systems the escape over the potential barrier out of the
attraction domain of one of the two minima of the bistable potential
is driven by an external additive or multiplicative DMN.
Consider the interval $[x_A, x_B]$ around a stable state $x_s$ of $V(x)$
(i.e., $f(x_s)=0,\, f'(x_s)<0$), containing a potential barrier (an unstable state) of $V(x)$ at $x_u$ (i.e., $f(x_u)=0,\, f'(x_u)>0$).
Let us compute
the MFPTs $T_{\pm}(x_0)$ out of this interval, as  functions of the initial
position $x_0$. Let us admit that $x_A$ is a perfectly reflecting boundary
\begin{equation}
T_{-}(x_A)=T_+(x_A)
\end{equation}
and $x_B$ is an absorbing boundary
\begin{equation}
T_+(x_B)=0\,,
\end{equation}
and, moreover, that the two alternate flows $f_{+}(x)=f(x)+Ag(x)$ and $f_{-}(x)=f(x)-Ag(x)$
have opposite signs throughout $[x_A, x_B]$ (for example, $f_+(x)>0$ and $f_-(x)<0$,
$\forall x\in [x_A, x_B]$).
Then one finds (see Ref.~\cite{balakrishnan88II}):
\begin{eqnarray}
&&T_+(x_0)=\nonumber\\
&&\int_{x_0}^{x_B} dz\;\frac{g(z)}{D[g(z)+f(z)/A]^2[g(z)-f(z)/A]P_{st}(z)}
\int_{x_A}^z du \;\frac{[g(u)+f(u)/A]P_{st}(u)}{g(u)}\nonumber\\
&&\nonumber\\
&&+\frac{DP_{st}(x_A)[g^2(x_A)-f^2(x_A)/A^2]}{Ag(x_A)}\int_{x_0}^{x_B}du\;
 \frac{g(u)}{D[g(u)+f(u)/A]^2[g(u)-f(u)/A]P_{st}(u)}\,,\nonumber\\
 \label{tp}
\end{eqnarray}
and
\begin{eqnarray}
&&T_-(x_0)=T_-(x_B)\nonumber\\
&&+\int_{x_0}^{x_B} dz\;\frac{g(z)}{D[g(z)-f(z)/A]^2[g(z)+f(z)/A]P_{st}(z)}
\int_{x_A}^z du \;\frac{[g(u)-f(u)/A]P_{st}(u)}{g(u)}\nonumber\\
&&\nonumber\\
&&-\frac{DP_{st}(x_A)[g^2(x_A)-f^2(x_A)/A^2]}{Ag(x_A)}\int_{x_0}^{x_B}du\;
 \frac{g(u)}{D[g(u)+f(u)/A]^2[g(u)-f(u)/A]P_{st}(u)}\,.\nonumber\\
 \label{tm}
\end{eqnarray}
The value of $T_-(x_B)$ can be obtained by taking $x_0=x_A$ in equation~(\ref{tm})
and using the perfectly reflecting boundary condition $T_-(x_A)=T_{+}(x_A)$,
together with equation~(\ref{tp}) for $x_0= x_A$.

Here $P_{st}(x)$ is the stationary probability density of the dichotomous flow,
\begin{equation}
P_{st}(x)=\frac{g(x)}{[g(x)+f(x)/A][g(x)-f(x)/A]}\,\int_0^x dz\;
\frac{f(z)}{D[g(z)-f(z)/A][g(z)+f(z)/A]}\,,
\end{equation}
and $D=A^2/2k$.

In the weak-noise limit $D \ll 1$ ( for $x_A<x_s<x_u$ and 
 $x_B$ being the other stable point of the potential $V(x)$),
using the steepest descent approximation, one obtains
for the activation rate over the barrier at $x_u$:
\begin{equation}
{\cal {R}}=\frac{1}{T_+}=\frac{\sqrt{|f'(x_s)|\,f'(x_u)}}{2\pi}\;\exp(-\Delta \Phi/D)\,,
\end{equation}
independent of $x_0$ (at order ${\cal O}(D^0)$), 
and with an ``Arrhenius barrier"
\begin{equation}
\Delta \Phi=-\int_{x_s}^{x_u} du\;\frac{f(u)}{[g(u)-f(u)/A][g(u)+f(u)/A]}\,.
\end{equation}

 One can also show that, up to order ${\cal O}(D^0)$,
 the activation rate ${\cal {R}}$ is  equal
to the  Kramer's escape rate out of this basin of attraction.
The latter one can be evaluated  as the constant net flux of probability (or the flow of particles in an ensemble representation) through the
borders of the domain over the integrated probability (the population)
inside the domain (the so-called ``flux over population" method),
see also Refs.~\cite{hanggi90,reimann96}.
Thus Kramer's rate can be simply obtained from the stationary master equation for
the probability density of the process.

It has been  recently
shown in Ref.~\cite{reimann99},  that the result
written symbolically as~\footnote{See Ref.~\cite{reimann99} for the right interpretation of Eq.~(\ref{symbol})}:
\begin{equation}
\mbox{MFPT}=\frac{1}{\mbox{Kramer's escape rate}}
\label{symbol}
\end{equation}
holds actually for {\em arbitrary DMN strength}, and not only in the weak-noise limit.
And even more: it is valid for an arbitrary time-homogeneous (stationary) stochastic process, and not only for the DMN.
On the basis of this result, one can therefore express
the MFPT in terms of the stationary probability density
and avoid the complications
related to the use of the backward equation
with the cumbersome boundary conditions.

\subsection{Resonant activation over a fluctuating barrier}

A related problem, that is ubiquitous  in natural sciences,
is that of the thermally-activated escape over a
potential barrier. In many situations of interest the noise is additive
GWN and weak, and the escape time is governed by a simple Arrhenius factor
determined by the height of the barrier.

However,  far from equilibrium the interplay between the noise
and the global properties of the potential may be much more intricate.
In particular, the potential that is experienced by the Brownian particle~\footnote{Recall that the term ``Brownian particle" may refer to a true particle, but it may also represent  some other state variable or collective coordinate of the current system under study.} cannot be always  appropriately modeled as a static one, but it may be randomly
varying on a time scale that is comparable to the escape time itself;
this may have nontrivial resonant-like effects on the behavior of the particle.
A few relevant experimental realizations in physical, chemical, and biological systems  are presented in Ref.~\cite{reimann97} and cited references therein; they include glasses, dye lasers, biomolecular  ionic channel kinetics, protein folding, etc.

As an example, let an overdamped particle
which moves in a fluctuating potential $V(x,t)$
with a barrier,
under the influence of a heat bath at a temperature $T$, as described by the
Langevin equation
\begin{equation}
\dot{x}(t)=-V'(x,t)+\xi_{\mbox{GWN}}(t)\,,
\end{equation}
with $\langle \xi_{GWN}(t_1)\xi_{GWN}(t_2)\rangle =2T\delta(t_1-t_2)$
(provided that  particle's mass, the friction coefficient, and Boltzmann's constant $k_B$
were set equal to 1, and time is measured in terms of the friction coefficient).
Consider that $V(x,t)$ switches at random, {\em dichotomously}, with transition rate $k$, between two profiles, $V_{\pm}(x)$, with two values of the barrier height, $V_{-}^{\mbox{barrier}}<V_{+}^{\mbox{barrier}}$.
For very slow barrier fluctuation rates (significantly slower than the  time required
to cross the highest barrier), the MFPT, as expected, approaches the average
of the MFPTs for the alternate barrier configurations. Also, for barrier fluctuations that are fast compared to the typical crossing time, the MFPT approaches the value required to cross the average barrier $V_{\mbox{mean}}^{\mbox{barrier}}=(V_{-}^{\mbox{barrier}}+V_{+}^{\mbox{barrier}})/2$.
However, for {\em intermediate fluctuation rates}, a resonance-like phenomena was shown to occur,
namely, it was shown that the MFPT  over the barrier
goes through a minimum when  the correlation time
(the inverse of the switching rate)
of the fluctuating potential is of the order
of the escape time over the lowest barrier $V_{-}^{\mbox{barrier}}$.
More precisely, this minimum of the MFPT (as a function of $k$)
corresponds to a maximum in the probability that the barrier is in the ``down" configuration
at the instant of crossing, as a function of the switching rate $k$.

This  phenomenon was called {\em resonant activation}. Its various aspects and
different realizations, were approached theoretically in Refs.~\cite{doering92,vandenbroeck93,bier93,zurcher93,pechukas94,marchesoni95,reimann96,marchi96,gaveau96,schmid99,li99,lehmann00}, including  comparisons with a deterministic,
periodic oscillation of the barrier in Ref.~\cite{jung93,klik01}. A simple experimental realization in an electronic circuit was studied in Ref.~\cite{mantegna00}.
Comprehensive discussions of the  ``generic character" of the resonant activation
(i.e., its occurence for general forms
of the potential barriers, for various intensities of the GWN, as well as for other types of
correlated fluctuations of the potential barrier -- as, for example, fluctuations that are driven by an Ornstein-Uhlenbeck process) were given in Refs.~\cite{reimann95,reimann97,reimann97I}.

\begin{figure}[h!]
\quad \vspace{1.5cm} \\
\centerline{\epsfxsize=11.cm\epsfbox{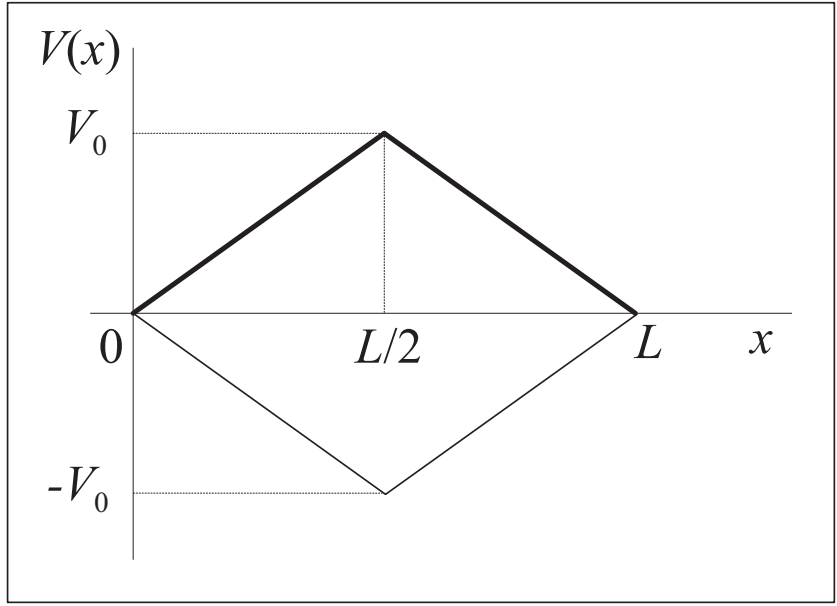}}
\caption{Piecewise linear symmetric potential that fluctuates dichotomously between
the configurations $V(x)$ and $-V(x)$, with the barriers $V_{+}^{\mbox{barrier}}=-V_{-}^{\mbox{barrier}}=V_0>0$.}
\label{figure22}
\end{figure}

Probably the simplest example, that admits a complete analytical treatment, can be found
in Ref.~\cite{doering92}, and corresponds to a symmetric piecewise potential that fluctuates, with a transition rate $k$, between $V(x)$ and $-V(x)$, with $V_{+}^{\mbox{barrier}}=-V_{-}^{\mbox{barrier}}=V_0>0$, see Fig.~\ref{figure22}.

\begin{figure}[h!]
\quad \vspace{1.5cm} \\
\centerline{\epsfxsize=11.cm\epsfbox{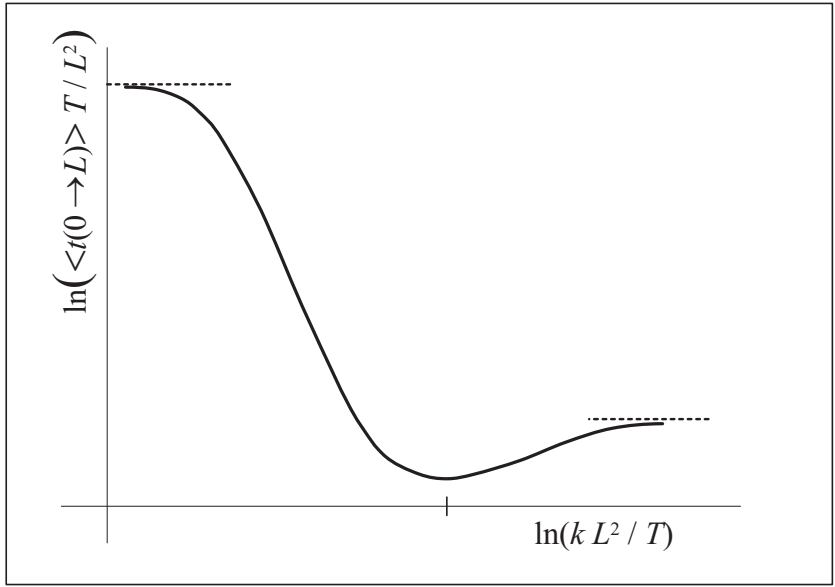}}
\caption{Schematic representation (continuous thick line) of the log of the dimensionless MFPT $\langle t(0\rightarrow L)\rangle$ as a function of the log of the dimensionless transition rate $k$ for the piecewise linear symmetric potential discussed in the main text, see Fig.~\ref{figure22}. The  dashed lines represent the two deterministic limits corresponding, respectively, to very slow and very high transition rates $k$.}
\label{figure23}
\end{figure}

The MFPT from $0$ to $L$
(provided that the particle starts in $x=0$, with the barrier in the $\pm V(x)$ configuration
with equal probability $1/2$, and it is absorbed at $x=L$)
is found to be:
\begin{eqnarray}
\frac{\langle t(0\rightarrow L)\rangle\; T}{L^2}=&&C_+\left[\frac{V_0}{\alpha T}(1-e^{-\alpha})-
\frac{kL^2\alpha}{V_0}-\frac{V_0}{T}\right]\nonumber\\
&&+C_-\left[-\frac{V_0}{\alpha T}(1-e^{\alpha})+
\frac{kL^2\alpha}{V_0}-\frac{V_0}{T}\right]\,,
\end{eqnarray}
where
\begin{equation}
\alpha=\left(\frac{V_0^2}{T^2}+\frac{2kL^2}{T}\right)^{1/2}
\end{equation}
and
\begin{equation}
C_{\pm}=\frac{(2kL^2/T)(\alpha\mp e^{\pm \alpha})\mp V_0^2/T^2}
{2\alpha V_0/T(1+2kL^2T/V_0^2)(2kL^2/T\mbox{cosh}\alpha +V_0^2/T^2)}\,.
\end{equation}
A qualitative representation of the MFPT as a function of the switching rate $k$
is given in Fig.~\ref{figure23}
for a fixed values of $V_0/T$. One notices the presence of a {\em minimum}
of the MFPT corresponding to a
certain value of the barrier fluctuation rate, that is found to be of the order of the
inverse of the time required to ``cross" the down potential configuration.
For the limit of slow transition rates,
one finds
\begin{equation}
\frac{\langle t(0\rightarrow L)\rangle_{\mbox{slow}}\;T}{L^2}=\frac{T^2}{2V_0^2}
\left(e^{V_0/T}-1-V_0/T\right)
+\frac{T^2}{2V_0^2}
\left(e^{-V_0/T}-1+V_0/T\right)
\end{equation}
(corresponding to the mean of MFPTs over the independent deterministic ``up" and ``down" potential configurations),
while in the limit of fast transition rates one finds the MFPT corresponding to the
transition over the mean barrier $\langle V_0\rangle=[V_0+(-V_0)]/2=0$ (i.e., the free diffusion on a line),
\begin{equation}
\frac{\langle t(0\rightarrow L)\rangle_{\mbox{fast}}\;T}{L^2}=
\lim_{\langle V_0\rangle \rightarrow 0} \frac{T^2}{\langle V_0\rangle ^2}
\left(e^{\langle V_0\rangle /T}-1-\frac{\langle V_0\rangle}{T}\right)=\frac{1}{2}\,.
\end{equation}

We also mention the recent results of Refs.~\cite{spa1,spa4,spa5} in which the
problem of escape time of a particle from a metastable state with a 
fluctuating barrier,
in the presence of both thermal noise and dichotomous noise, was solved in full
analytical detail. The related problem of steady-state distributions and the
exact results for the diffusion spectra were addressed in Refs.~\cite{spa2,spa3}. 
Finally, 
general equations for computing the effective diffusion coefficient in randomly
switching potentials are derived for arbitrary mean rate of the potential
switching and arbitrary intensity of the Gaussian white noise,
Refs.~\cite{spa5,spa6,spa7}.

\section{STOCHASTIC RESONANCE WITH DMN}

Generally speaking, the {\em stochastic resonance} (SR) phenomenon
refers to the ``enhanced sensitivity" of a
(nonlinear) system to a small deterministic periodic forcing
in the presence of an ``optimal amount" of noise -- see Ref.~\cite{gammaitoni98}
for a (by now incomplete) review of various model-systems with  SR and their possible practical applications.
Such a definition is very broad, and till now there is no
agreement about  the precise signature of the SR, the necessary conditions of its occurrence,
as well as the ``right" quantifiers, see Refs.~\cite{berdi98,evsti05}
for further comments on this point. There is therefore a huge and varied literature on SR and its numerous realizations.  In particular, for systems with DMN-driving see Refs.~\cite{berdichevsky96,barzykin98,berdichevsky99,neiman99,rozenfeld00,freund00,gitterman03} for a few examples.

\begin{figure}[h!]
\quad \vspace{1.5cm} \\
\centerline{\epsfxsize=9.cm\epsfbox{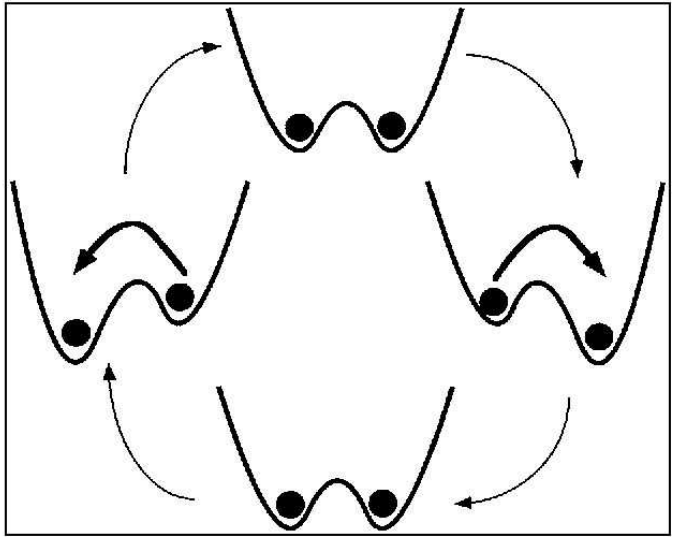}}
\caption{Schematic representation of a double-well symmetric  potential
that is periodically rocked by a weak external signal $s(t)$.
Considered are four stages of one rocking period, corresponding 
to successive maximum and minimum heights of the potential barrier.
For an intermediate, optimum thermal noise intensity, 
there is a synchronization between the
rocking period and the thermally-activated hopping 
between the two wells.}
\label{figure24}
\end{figure}

A canonical model for SR is an overdamped particle in a symmetric double-well potential
$V(x)=-ax^2+bx^4$ (with $a,\, b>0$), driven simultaneously by an additive GWN and an additive, weak periodic signal $s(t)$, see Fig.~\ref{figure24} for an illustration. A ``hand-waving" argumentation for the appearance of SR would be the following. One one side,
for too low GWN intensities, the thermally-activated jumps between the two wells are too rare and the particles do not take benefit of the alternate decrease of the potential barrier (on one side or another) due to the external signal; on the other side, for too large GWN intensities, the jumps between the wells are very frequent (a large number take place during one period of the external signal) and thus,
again, the response of the system is not synchronized with and does not benefit of the external signal.
However, for intermediate noise intensities, the thermally-activated transition rates are comparable with the rocking rate
of the potential, and the particles take advantage of the alternate decrease of the potential barrier,
resulting in an enhanced response of the system to the applied external perturbation.
As shown, for example, in Ref.~\cite{gammaitoni98},
both the spectral power amplification (SPA) (that represent the weight of the signal part in the output power spectrum), and the signal-to-noise ratio (SNR) (the SPA rescaled by the input power) represent good measures of the SR. Indeed, both of them show a {\em nonmonotonous behavior
with a  maximum as a function of the GWN intensity}.

As it was recently shown, see Refs.~\cite{neiman99,rozenfeld00,freund00}, the addition of a  DMN
has very important effects on the behavior of the system: (i) DMN can synchronize the switching time between the two stable states of the double-well (i.e., for a certain interval of the GWN intensity, the mean switching rate of the system
is locked to the switching rate of the DMN), a phenomenon corresponding,
in the limit of a weak external perturbation $s(t)$,
to the {\em resonant activation} described in Sec.~VI.B; (ii) moreover, the {\em SR  is greatly enhanced by the DMN} (i.e., the SPA and/or the SNR can reach larger maximal values as compared to the case when no DMN is present)~\footnote{Note that the DMN is {\em weak} and cannot, by itself, induce transitions between the stable states of the double-well potential: the jumps are still  induced by the GWN.}.

Following Refs.~\cite{neiman99,rozenfeld00,freund00}, let us illustrate these results on
a simplified model.  We shall neglect the intrawell motion, and
describe only the belonging of the particle to one or the other
wells of $V(x)$ through the two-valued stochastic variable $\sigma(t)=\pm 1$.
The thermally-induced transition rate between the two-wells is given by
$a=\exp(-\Delta V/D)$, where  $\Delta V$ is  the height of the potential barrier,
and $D$ is the properly-scaled GWN intensity.
An additive DMN $\xi(t)={\pm A}$ of low switching rate $k\ll 1$ modifies these transition rates,
\begin{equation}
W(\sigma, \xi)=\exp\left(-\frac{\Delta V+\sigma \xi}{D}\right)\,,\quad\sigma=\pm 1\,,\xi=\pm A\,.
\label{trate}
\end{equation}
Considering the four-state stochastic process $(\sigma,\,\xi)$
one can write down the following master equation:
\begin{equation}
\frac{d}{dt}P(t; \sigma,\xi)=-W(\sigma,\xi)P(t;\sigma,\xi)+W(-\sigma,\xi)P(t;-\sigma,\xi)+k[P(t;\sigma,-\xi)-P(t;\sigma,\,\xi)]\,.
\label{MSRE}
\end{equation}
Therefore, the mean switching rate (MSR) of the output $\sigma(t)$ is found to be:
\begin{equation}
{\cal {R}} =\frac{\pi}{2}\left(a_1+a_2-\frac{(a_2-a_1)^2}{a_1+a_2+2k}\right)\,,
\end{equation}
with $a_{1,2}=\exp[-(\Delta V\pm A)/D]$. For small values of the GWN intensity
$D\ll1$, one recovers the limit of fast switching rates of the DMN, $k\gg 1$,
that corresponds to an {\em effective potential with a lowered barrier $\Delta V-A$}:
\begin{equation}
{\cal {R}} \approx \pi/2 \exp[-(\Delta V-A)/D] \quad (D\ll 1)\,.
\end{equation}
 For  large noise intensities, the MSR approaches the other deterministic limit of the DMN, namely $k\rightarrow 0$, that corresponds to {\em a double-well static potential with an  asymmetry},
\begin{equation}
{\cal {R}} \approx \pi \exp(-\Delta V)/\mbox{cosh}(A/D) \quad (D>1)\,.
\end{equation}
For intermediate values of $D$ (and finite values of $k$), the MSR is
locked to $\approx k$, and changes very slowly with $D$, a regime that corresponds to the
{\em resonant activation} described in Sec.~VI.B above.

Let us now add a small periodic perturbation to the system, $s(t)=S\cos(\Omega t+\varphi)$,
with $\varphi$ randomly uniformly distributed in $[0,2\pi)$. At slow
($\Omega <k\ll 1$) and weak ($S\ll A< \Delta V $) forcing, the transition rates in equation~(\ref{trate}) become
\begin{equation}
W(\sigma,\xi) \rightarrow W(\sigma,\xi)\exp[-S\sigma \cos(\Omega t+\varphi)/D]\approx
W(\sigma,\xi)[1-S\sigma \cos(\Omega t+\varphi)/D]\,,
\label{modtrate}
\end{equation}
and the evolution of the system is described by the master equation~(\ref{MSRE})
with the time-periodic transition rates~(\ref{modtrate}). The autocorrelation function
$\langle \sigma(t_1)\sigma(t_2)\rangle$ can be computed in closed analytical form
in the limit of the {\em linear-response theory} with respect to the signal $s(t)$. This leads finally to the following output power spectrum (averaged over the phase $\varphi$ of the external
periodic signal $s(t)$):
\begin{equation}
{\cal{P}}(\omega)={\cal{N}}(\omega)+S^2\pi\eta\delta(\omega-\Omega)\,,
\end{equation}
 where
 \begin{equation}
{\cal{N}}(\omega) =\frac{4(a_1+a_2)}{(a_1+a_2)^2+\omega^2}\left[1+\frac{(a_2-a_1)^2}{4k^2+\omega^2}\right]-\frac{4(a_2-a_1)^2}{(a_1+a_2+2k)(4k^2+\omega^2)}
\end{equation}
is the power spectrum of the background (independent of the applied periodic
perturbation $s(t)$). The SPA is found to be:
\begin{equation}
\eta=\frac{4}{\pi^2D^2}\;\frac{{\cal R}}{(a_1+a_2)^2+\Omega^2}\,,
\end{equation}
and the signal-to-noise ratio is:
\begin{equation}
SNR=\pi \frac{\eta}{{\cal{N}}(\Omega)}\,.
\end{equation}
Both the SPA $\eta$ and the  SNR exhibit two maxima as a function of the GWN intensity,
one for low noise intensities, and the other one, weaker, for large noise intensities.
As seen above, the small-noise intensity region  $D\ll 1$ corresponds to a potential with a lowered barrier, so that SR may occur for smaller noise intensity as compared to the case
when no DMN is present: that is why SR is greatly enhanced. In the intermediate range of $D$,
the MSR is locked to the switching rate of the DMN, and thus the system is not sensitive to the periodic perturbation: SR does not appear. Finally, for large values of the noise intensity, the system behaves as an asymmetric one: SR still appears (the second peak), but during one-round trip switching of the DMN the system performs many transitions between the potential wells; thus, with further increase of the asymmetry (of the noise-intensity $D$)
the SR is gradually suppressed.

This discussion highlights the {\em relationship between the} {\em resonant activation over a fluctuating barrier} (see Sec.~VI.B) and the {\em (enhanced) stochastic resonance}: although  both phenomena may appear in the same system, they are typically  appearing in quite different regimes of the parameters of the system. This conclusion is corroborated, in a slightly different context,  by the results of Ref.~\cite{schmitt06}.

\section{SPATIAL PATTERNS INDUCED BY DMN}

Till now, with the exception of noise-induced
phase-transition situations,
we considered only zero-dimensional systems, i.e., systems
with no spatial dimensions.
We will now briefly turn to the spatially-extended systems,
for which it was shown that noise,
and in particular DMN, can create {\em spatial patterns},
see Refs.~\cite{parrondo96,ojalvo99,buceta02I,buceta02II,buceta02III,buceta02IV,buceta03I,buceta03II,buceta03III,parrondo03,wio03}.

Pattern formation in a non-fluctuating system is generically related to
the onset of a symmetry-breaking global instability of
the homogeneous state of the system. It appears
when the control parameter
(that is related to the ratio between the driving forces
-- that tend to destabilize the system -- and
the dissipative forces in the system --
that have a stabilizing effect) exceeds a
threshold value, see e.g. Ref.~\cite{cross77}. In the vicinity of the
instability threshold, the dynamics of the system is essentially described by
{\em nonlinear amplitude equations}, that correspond
to the evolution of the relevant slow modes on
the central manifold of the critical point
of the dynamics. These equations are generic, in the sense that
their form does not depend on the details of the underlying dynamics, but
only on the type of bifurcation that appears in the system,
the symmetries of the system, the dimensionnality, and the eventual
conservation laws.

A well-known example of amplitude equation is the
Swift-Hohenberg equation, that was succesfully used to describe
the onset of the Rayleigh-B\'enard convective rolls. It corresponds to a system
undergoing a soft pitchfork bifurcation,
described by a single scalar field $\psi(\vec{r},\,t)$, provided that
the system is invariant under spatial inversion $\vec{r} \rightarrow -\vec{r}$
and under field-inversion $\psi \rightarrow -\psi$. Its form is:
\begin{equation}
\partial_t
\psi(\vec{r},t)=-V'(\psi(\vec{r},t))-(K_c^2+\Delta^2)^2\psi(\vec{r},t)\,,
\end{equation}
where $\Delta$ is the Laplace operator, and $V(\psi)$ is an even function of $\psi$
(in the simpest case, $V(\psi)$ is just a quartic, symmetric potential),
with $V'(\psi)=dV/d\psi$.
As long as $V(\psi)$ is monostable, no spatial structures appear in the system --
the stable steady-state is spatially homogeneous. However, if under the variation of the control parameter  the local
potential $V(\psi)$  acquires two stable equilibrium points (separated by a barrier),
the Swift-Hohenberg model leads to pattern formation with a critical wavevector $K_c$. More precisely, one has the following conditions for the marginal stability:
\begin{equation}
V'(\psi)+K_c^2\psi=0\,,\quad V''(\psi)<0\,.
\label{marstab}
\end{equation}

In the case of {\em noise-induced spatial patterns}, the appearance of the pattern is
strictly conditioned by the presence of the noise, i.e., no pattern appears in the
noiseless, deterministic system. Such an example is offered by the
Swift-Hohenberg equation with a time-dependent potential $V(\psi,t)$
that switches at random,
{\em dichotomously}, between two profiles, $V_1(\psi)$ and $V_2(\psi)$
(a so-called dichotomous global alternation of dynamics), see
Refs.~\cite{buceta02II,buceta02III,buceta02IV}. We consider that both
potentials
$V_{1,2}(\psi)$ are monostable, i.e., neither of the two dynamics alone will
lead to patterns (there are no patterns in the absence of the DMN).

However, it
was shown that the {\em random alternation} of the two dynamics leads to
the appearance of {\em stationary spatial patterns}, depending on the characteristics of
the potentials
$V_{1,2}$ and on the switching rate $k$ of the DMN. This result can be put in
parallel with the well-known Parrondo paradox (or the flashing ratchet mechanism),
in which the (random) alternation between two fair games (or two unbiased
diffusions) is no longer fair (is no longer unbiased);
this means that this alternation  leads to a
directed flow in the system -- i.e., a net gain (a directed current). See
Refs.~\cite{parrondo03,wio03} for a pedagogical discussion of these points.

A simple qualitative argument can clarify the underlying mechanism. Let $\tau=1/k$
the average time the system is spending in each of the alternate dynamics.
Of course, if the switching rate is very low, $k\rightarrow 0$, in-between the switches
the field will reach, respectively, the stationary homogeneous states $\psi_{1,2}$
corresponding to each of the monostable potentials $V_{1,2}(\psi)$:
\begin{equation}
V_{1,2}(\psi_{1,2})+K_c^2\psi_{1,2}=0\,,\quad V''_{1,2}(\psi_{1,2})<0\,.
\end{equation}

Suppose now that the switching rate is very high (the sojourn time $\tau$ is much smaller than the characteristic relaxation time in each of the alternate $V_{1,2}$ potentials). Then the evolution of the field will be driven by the deterministic {\em effective potential} $V_{\mbox{eff}}(\psi)=[V_1(\psi)+V_2(\psi)]/2$.
If
\begin{equation}
V_{\mbox{eff}}({\psi})+K_c^2\psi=0 \quad \mbox{and}\quad V''_{\mbox{eff}}({\psi})>0\,,
\label{effective}
\end{equation}
then, according to the general considerations above, the system becomes marginally unstable with respect to the onset of patterns of wavevector  $K_c$. Note that $V_{\mbox{eff}}$ may become bistable
only for {\em nonlinear} systems (i.e., for potentials $V_{1,2}$ that are at least quadratic).
Of course, by continuity arguments one may expect
the onset of the spatial patterns for finite switching rates $k$, too, 
but the marginal stability condition Eq.~(\ref{effective}) will become dependent on $k$.

When the global
switching is not random, but periodic in time, see
Ref.~\cite{buceta02I,buceta02IV}, besides the stationary spatial patterns one will
also obtain periodic spatio-temporal patterns.
The case of a potential $V(\psi, \vec{r})$ that has
a spatial, quenched dichotomous disorder was also shown to lead to spatial
patterns, see Ref.~\cite{buceta03I}.

Moreover, as pointed out in
Refs.~\cite{buceta02III,buceta03III}, these studies are relevant
for other situations that lead to pattern formation, such as Turing
instabilities in reaction-diffusion systems.

\section{DISCRETE-TIME DICHOTOMOUS FLOWS}

The influence of noise on {\em discrete-time} dynamical systems
(maps) is much less-documented than for the continuos-time ones.
Aspects like the shift, broadening, or even suppression of bifurcations,
the behavior of the invariant densities and of the
Lyapunov exponents near the onset of chaotic behavior,
and the destabilization of locally stable states by noise
have been documented, in general, for weak GWN, see Refs.~\cite{cru81,shraiman81,hirsch82,arecchi84,linz86,beale89,reimann91,graham91,reimann94} and references therein for examples.

With few exceptions, see Refs.~\cite{irwin90,fraser92,gutierrez93,kosinka03},
the effects of a finite correlation time of the noise,
such as a DMN, have not been addressed in general.
For example,  Ref.~\cite{gutierrez93} considers
the logistic map with a {\em dichotomously fluctuating parameter}:
\begin{equation}
x_{n+1}=\mu(1+\xi_n)x_n(1-x_n)\,.
\end{equation}
$\xi_n$ is a dichotomous noise $\pm A$ with a probability $\alpha$
of repeating the same value in the next iteration
($\alpha=1/2$ corresponds to the white noise limit,
while $\alpha=0$ and $\alpha=1$ correspond to two deterministic limits),
and $\mu(1+A)<4$.
Such a system may derive from a continuous-time one which is
driven by a random sequence of pulses of constant duration
(this duration thus constitutes the time step of the map).
The influence  of the
``correlation time" $\alpha$
of the noise on the dynamics  is  found to be quite dramatic:
by simply varying it one can obtain all transitions,
from chaos to regular motion.

However, this field definitely calls for further investigations.

\section{CONCLUSIONS AND PERSPECTIVES}

The main goal of this review was to present the DMN as a flexible and very efficient tool in modeling out-of-equilibrium situations,  allowing to emphasize the richness and variety of the encountered phenomena. We hope that the reader will find in it an useful ingredient in his/her
modeling of various experimental situations, out of which only a small part was reported in this paper.
In this respect, we are thinking of the cutting-edge techniques that allow to go to
the direct study and visualization of
microscopic and nanoscopic systems (including biological,
living systems): we expect coloured noise (and, in particular, DMN)
to play an important role in modeling the interaction of  such small systems with their surroundings.

We also emphasize a few other opened conceptual problems. A first one would be the effect
of  inertia on dichotomous flows, and, more general, the statistical properties of
several coupled dichotomous flows. The problem of fluctuation-like theorems for DMN-driven systems is to be addressed in further details.
Also, as it was mentioned above, the study of maps driven by DMN is still an opened field.

\section*{Acknowledgements}

We are grateful to Profs. Christian Van den Broeck, Michel Droz,
Katja Lindenberg,  Ryoichi Kawai,
Venkataraman Balakrishnan, Avadh Saxena, Mamad Malek Mansour, Florence Baras,
and Max-Olivier Hongler
for over-the-years discussions on DMN-related subjects.
We also thank Fran\c{c}ois Coppex for precious help in
preparing the manuscript and acknowledge partial support from the Swiss
National Science Foundation.


\end{document}